\documentclass[twocolumn]{aastex631}

\usepackage{amsmath,gensymb}
\usepackage{xcolor}
\usepackage{mathrsfs}
\usepackage{booktabs,multirow}

\newcommand{\new}[1]{#1}

\begin{document}

\title{Supersensitive seismic magnetometry of white dwarfs}

\author[0000-0002-1884-3992]{Nicholas Z. Rui}
\affiliation{TAPIR, California Institute of Technology, Pasadena, CA 91125, USA}

\author[0000-0002-4544-0750]{Jim Fuller}
\affiliation{TAPIR, California Institute of Technology, Pasadena, CA 91125, USA}

\author[0000-0001-5941-2286]{J. J. Hermes}
\affiliation{Department of Astronomy, Boston University, 725 Commonwealth Ave., Boston, MA 02215, USA}

\begin{abstract}
The origin of magnetic fields in white dwarfs (WDs) remains mysterious.
Magnetic WDs are traditionally associated with field strengths $\gtrsim1\,\mathrm{MG}$, set by the sensitivity of typical spectroscopic magnetic field measurements.
Informed by recent developments in red giant magnetoasteroseismology, we revisit the use of WD pulsations as a seismic magnetometer.
WD pulsations primarily probe near-surface magnetic fields, whose effect on oscillation mode frequencies is to asymmetrize rotational multiplets and, if strong enough, suppress gravity-mode propagation altogether.
The sensitivity of seismology to magnetic fields increases strongly with mode period and decreases quickly with the depth of the partial ionization-driven surface convective zone.
We place upper limits for magnetic fields in $24$ pulsating WDs: $20$ hydrogen-atmosphere (DAV) and three helium-atmosphere (DBV) carbon--oxygen WDs, and one extremely low-mass (helium-core) pulsator.
These bounds are typically $\sim1$--$10\,\mathrm{kG}$, although they can reach down to $\sim10$--$100\,\mathrm{G}$ for DAVs and helium-core WDs in which lower-frequency modes are excited.
Seismic magnetometry may enable new insights into the formation and evolution of WD magnetism.
\end{abstract}


\section{Introduction} \label{sec:intro}

\begin{figure}
    \centering
    \includegraphics[width=0.47\textwidth]{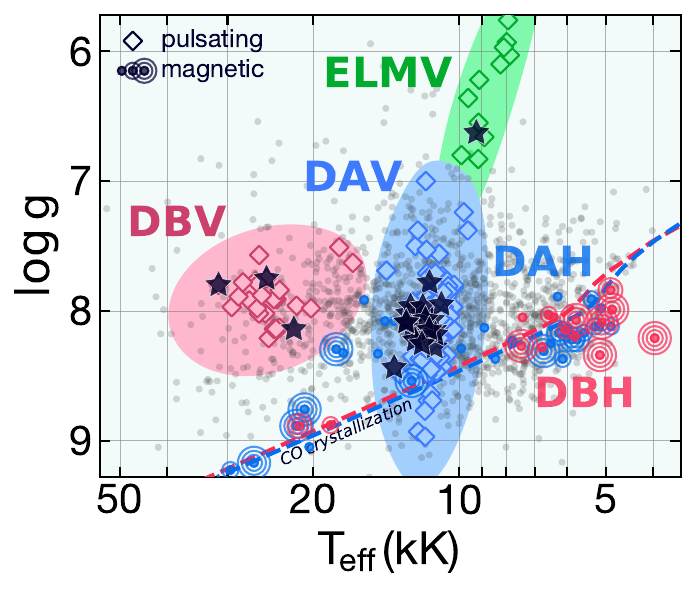}
    \caption{Kiel diagram showing various classes of pulsating \citep[\textit{diamonds};][]{corsico2019pulsating} and magnetic \citep[\textit{colored circles};][with some values of $\log g$ taken from the Gaia EDR3 catalog of \citealt{gentile2021catalogue}]{bagnulo2021new,bagnulo2022multiple} WDs.
    \textit{Black stars} indicate WDs whose magnetic fields we constrain in this study, while blue points are DA WDs and red points are DB WDs from the literature.
    Following the plotting convention of \citet{bagnulo2022multiple}, magnetic WD points surrounded by zero, one, two, and three \textit{concentric circles} have measured fields $B<1\,\mathrm{MG}$, $1\,\mathrm{MG}\leq B<10\,\mathrm{MG}$, and $10\,\mathrm{MG}\leq B<100\,\mathrm{MG}$, and $B\geq100\,\mathrm{MG}$, respectively.
    The background \textit{translucent gray points} indicate a subsample of other WDs in the catalog of \citet{gentile2021catalogue}.
    The innermost $5\%$ by mass of carbon--oxygen WDs with thin (thick) hydrogen envelopes will crystallize once they reach the \textit{red} (\textit{blue}) \textit{dashed lines} \citep{bedard2020spectral}.
    }
    \label{fig:wd_summary}
\end{figure}

White dwarfs (WDs) are the compact remnants of low- and intermediate-mass ($\lesssim8M_\odot$) stars.
Although a large fraction of WDs are now known to be magnetic \citep[$\approx20\%$;][]{bagnulo2021new}, the origins of their magnetic fields are still largely mysterious.

Magnetic fields in WDs are typically measured using Zeeman splitting of spectral absorption or emission lines \citep{landstreet2014basics,ferrario2015magnetic}.
Recent volume-limited surveys have revealed that magnetism in typical WDs (with masses $M\leq0.75M_\odot$) experience a delayed onset, with both the incidence and strength of magnetism increasing at cooling ages of $2$--$3\,\mathrm{Gyr}$ \citep{bagnulo2021new,bagnulo2022multiple}.
The late appearance of magnetic fields in these WDs may be due to some combination of an outward-diffusing fossil field from earlier evolutionary stages and a magnetic dynamo activated during core crystallization \citep{isern2017common,schreiber2021origin,ginzburg2022slow,blatman2024magnetic,fuentes2024short,blatman2024magnetic2}.
In contrast, strong magnetic fields occur in a large fraction ($\sim40\%$) of ultramassive ($M\gtrsim1.1M_\odot$) WDs \citep{bagnulo2022multiple,kilic2023merger}, suggesting a merger-related origin may be responsible for some of them \citep{tout2008binary,garcia2012double,briggs2018genesis,schneider2020long}.
The puzzle of WD magnetism is a timely one.

In parallel with these developments, leaps and bounds have been made in the asteroseismic inference of magnetic fields in the interiors of red giants in the last few years.
The propagation of buoyancy-restored gravity waves through stably stratified regions (such as in the radiative cores of red giants) is sensitive to the magnetic field.
The degree to which a standing gravity wave (g mode) of period $P$ is locally influenced by magnetism is determined by the comparison between the radial component of the magnetic field, $B_r$, to the critical field
\begin{equation} \label{bcrit}
    B_{r,\mathrm{crit}} \sim \frac{2\pi^2}{\sqrt{\ell(\ell+1)}}\frac{\sqrt{4\pi\rho}\,r}{NP^2}\mathrm{,}
\end{equation}

\noindent where $\rho$ is the density, $r$ is the radial coordinate, and $N$ is the Brunt--V\"ais\"al\"a (buoyancy) frequency, given in Gaussian units \citep{fuller2015asteroseismology}.

Magnetism (due to fields $\gtrsim100\,\mathrm{kG}$) is thought to dampen or outright suppress dipolar oscillations in approximately one-fifth of observed red giants (\citealt{garcia2014study,stello2016prevalence,cantiello2016asteroseismic}, although see \citealt{mosser2017dipole}).
This magnetic suppression can occur when $B_r>B_{r,\mathrm{crit}}$ somewhere in the g-mode cavity \citep{fuller2015asteroseismology,lecoanet2017conversion,lecoanet2022asteroseismic,rui2023gravity}.
However, in the last few years, asteroseismic frequency \textit{shifts} due to weaker magnetic fields of tens to a hundred kilogauss have been detected for the very first time \citep{li2022magnetic,deheuvels2023strong,li2023internal,hatt2024asteroseismic}.
Magnetic frequency shifts depend not only the magnetic field's strength but also its geometry.

The majority of existing red giant magnetic field measurements based on seismic frequency shifts have relied on asymmetries in observed dipole ($\ell=1$) triplets \citep{li2022magnetic,li2023internal}.
While rotation (to first order) splits a single peak in the asteroseismic power spectrum into a symmetric multiplet of $2\ell+1$ distinct modes, the Lorentz force typically causes asymmetric splitting within the multiplet \citep{bugnet2021magnetic,li2022magnetic,das2024unveiling}.
Even when other sources of asymmetry cannot be excluded \citep[e.g., near-degeneracy effects;][]{deheuvels2017near,ong2022mode}, the degree (or lack) of asymmetry imposes an upper bound on the magnetic field.

The concept of performing similar, asymmetry-based magnetic field measurements in pulsating WDs dates back to \citet{jones1989possibility}, who predicted that seismology would be sensitive to weak fields far below the ``traditional'' megagauss WD magnetic field.
Soon after, \citet{winget1994whole} reported the seismic detection of a kilogauss-level magnetic field in Whole Earth Telescope observations of the brightest DBV, GD 358 (i.e., the prototype DBV, V777 Herculis).
Similar seismic field constraints have been placed on various other WDs over the years \citep{kawaler1995whole,schmidt1997upper,vauclair2002asteroseismology,dolez2006whole,fu2007asteroseismology,hermes2017deep}, but a uniform analysis for a large sample of WDs has not been performed.
Although DQVs have previously been thought to be pulsating, magnetic WDs \citep{dufour2008sdss}, their variability is now generally believed to be due to surface spots \citep{williams2016variability}.
We are not aware of any definitive detections of pulsations in known magnetic WDs.

In this work, we use asteroseismic data to place approximate upper limits on the magnetic fields in the near-surface layers of $24$ WDs.
In doing so, we aim to add seismology to the toolkit of WD magnetometry.
Seismic magnetometry has the potential to complement traditional Zeeman effect-based techniques in constraining the formation and evolution of WD magnetic fields (Figure \ref{fig:wd_summary}).
Section \ref{theory} summarizes our present understanding of gravity waves under magnetic fields.
Section \ref{methods} describes our stellar models and the procedure by which we place seismic upper bounds on WD fields.
Section \ref{seismicconstraints} presents and discusses our findings.
Section \ref{summaryfuture} summarizes and presents optimistic prospects.

\section{Theoretical background} \label{theory}

Here, we outline the existing formalism on gravity waves under the influence of the Lorentz force.
Sections \ref{sectfrequencyshifts} and \ref{sectmodesuppression} summarize theoretical predictions for frequency shifts and conditions for mode suppression, respectively.

The following formulae for the frequency shifts are valid in the perturbative regime where both magnetism and rotation are treated at lowest order, with the additional assumption that rotation is strong enough to set the preferred direction of the problem \citep[see][]{li2022magnetic,das2024unveiling}.
Mode suppression conditions arise from a non-perturbative treatment of the magnetogravity problem \citep[e.g.,][]{fuller2015asteroseismology,lecoanet2017conversion,lecoanet2022asteroseismic,rui2023gravity,rui2024asteroseismic}.
Both calculations rely heavily on the incompressible approximation (which filters out pressure waves), the asymptotic approximation (which relies on high radial orders), and the Cowling approximation.
Departures from these assumptions are treated in this work as an ad hoc correction factor (Appendix \ref{sectnonasymptotic}).
Under these assumptions, g modes are primarily sensitive to the radial component of the magnetic field $B_r$ and insensitive to the horizontal components $B_\theta$ and $B_\phi$, which are hereafter neglected.

The aim of this work is to provide a constraint on WD field using both magnetic asymmetries and mode suppression.
While both effects are primarily sensitive to the outer layers of the WD (as we subsequently explain), the exact layers probed by each effect are different, and additionally they depend on the mode period, $P$.
Since all seismic methods we discuss probe the field in geometrically thin, near-surface layers, we speak of bounding ``the'' surface field $B_r$, which assumes that the magnetic field is roughly constant near the surface of the WD.

\begin{figure}
    \centering
    \includegraphics[width=0.48\textwidth]{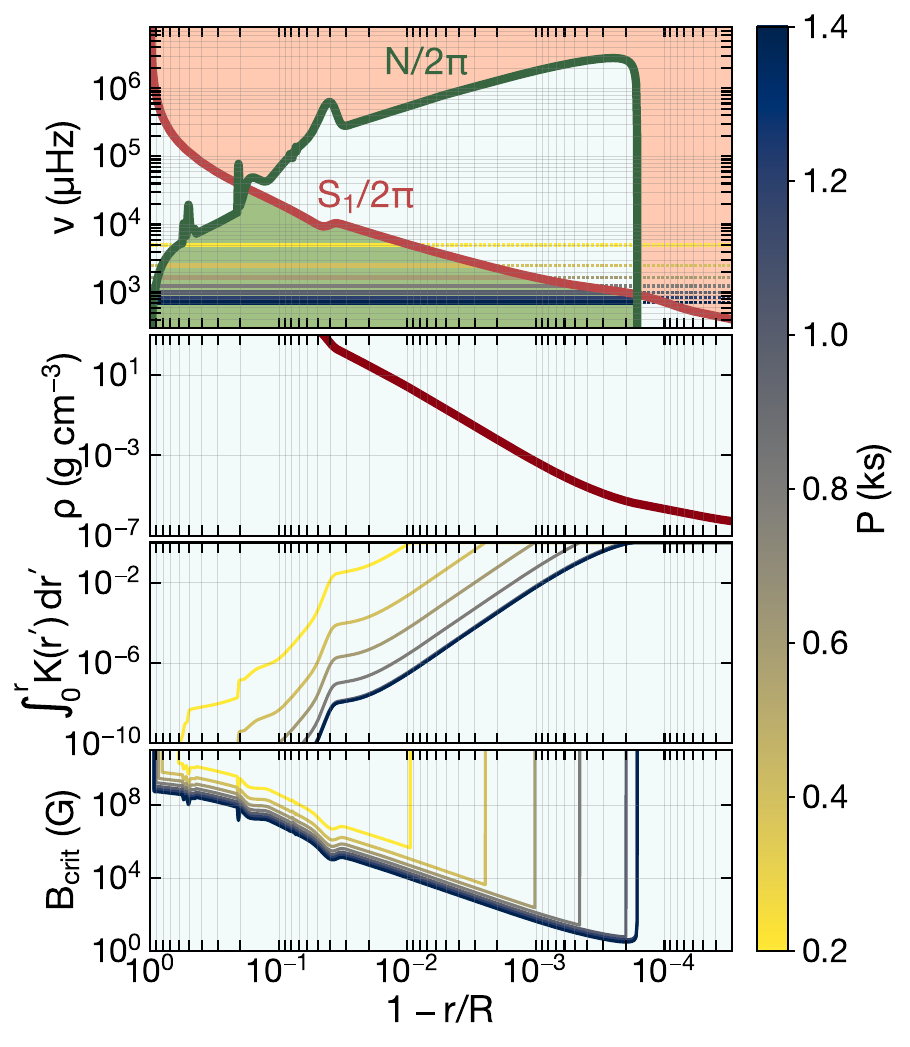}
    \caption{\textit{Top:} Asteroseismic propagation diagram for the dipole ($\ell=1$) modes of a WD model with $M=0.64M_\odot$ and $T_{\mathrm{eff}}=12\,\mathrm{kK}$, with $1-r/R$ on the $x$ axis (fractional radius from the outermost shell of the model).
    The \textit{green} and \textit{brown} curves show the Brunt--V\"aisala ($N$) and dipole Lamb ($S_1$) frequencies, respectively.
    The \textit{green shaded region} denotes the g-mode cavity.
    \textit{Second from top:} Density as a function of radius.
    \textit{Third from top:} Cumulative magnetic weight function (defined in Equation \ref{weightfunction}) for the same model, for a variety of mode periods.
    The function $\int^r_0K(r')\,\mathrm{d}r'$ represents the cumulative contribution of the magnetic field in shells within radii $r$ to the average $\langle B_r^2\rangle$ (Equation \ref{avgfield}).
    \textit{Bottom:} Critical field $B_{r,\mathrm{crit}}$ required to suppress gravity waves, for a variety of mode periods.
    }
    \label{fig:fancy_profile_main}
\end{figure}

\subsection{Seismic frequency shifts} \label{sectfrequencyshifts}

A non-rotating, non-magnetic star such as a WD possesses spherical symmetry, so that each radial order $k$ and angular degree $\ell$ are assigned to a multiplet of $2\ell+1$ modes of varying azimuthal order $-\ell\leq m\leq\ell$ with degenerate frequencies $\nu^{(0)}_{n\ell}$.
Due to geometric cancellation effects, observable g modes are typically low-$\ell$, either $\ell=1$ dipole modes or $\ell=2$ quadrupole modes, forming mode triplets and quintuplets, respectively.
The first-order effect of rotation is to split these modes into a symmetric multiplet \citep{ledoux1951nonradial,unno1979nonradial,aerts2010asteroseismology,aerts2021probing}:
\begin{equation}
    \nu_{k\ell m} \approx \nu^{(0)}_{k\ell} + m\left(1 - \frac{1}{\ell(\ell+1)}\right)\frac{\langle\Omega\rangle_g}{2\pi}\mathrm{,}
\end{equation}

\noindent where
\begin{equation}
    \langle\Omega\rangle_g \approx \frac{\int_{\mathcal{R}^\ell_{\;\nu}}\Omega\,(N/r)\,\mathrm{d}r}{\int_{\mathcal{R}^\ell_{\;\nu}}(N/r)\,\mathrm{d}r}
\end{equation}

\noindent is a wave-cavity-averaged rotation rate.
The symbol $\mathcal{R}^\ell_{\;\nu}$ denotes the g-mode cavity, the contiguous range of radii where the linear mode frequency $\nu$ satisfies $2\pi\nu<N,S_\ell$, where $S_\ell=\sqrt{\ell(\ell+1)}c_s/r$ is the degree-$\ell$ Lamb frequency \citep{unno1979nonradial,aerts2010asteroseismology}.
The rotational frequency shift is due to a combination of the Coriolis force (which is a physical effect) and a Doppler effect acting on the mode frequencies (which is a geometric one).

When the star is also magnetic, the magnetic tension serves as an additional, ``stiffening'' restorative force which shifts gravity mode frequencies by roughly

\begin{equation} \label{numag}
    \begin{split}
        \nu^\ell_B &\equiv \frac{A_\ell}{64\pi^5}\frac{P^3}{\int_{\mathcal{R}^\ell_{\;\nu}}(N/r)\,\mathrm{d}r}\int_{\mathcal{R}^\ell_{\;\nu}} \! \mathrm{d}r \frac{N^3}{\rho r^3} \int \frac{B_r^2}{4\pi} \, \mathrm{d}\Omega  \\
        &= A_\ell\frac{\mathscr{I}}{64\pi^5}\langle B_r^2\rangle P^3\mathrm{,}
    \end{split}
\end{equation}

\noindent where $\mathrm{d}\Omega$ is an integral over a spherical surface and
\begin{equation} \label{curlyIasympt}
    \mathscr{I} \simeq \frac{\int_{\mathcal{R}^\ell_{\;\nu}}(N^3/\rho r^3)\,\mathrm{d}r}{\int_{\mathcal{R}^\ell_{\;\nu}}(N/r)\,\mathrm{d}r}\mathrm{,}
\end{equation}

\noindent and $P=1/\nu$ is the period of the mode and $A_\ell=\ell(\ell+1)/2$.
The characteristic scale $\nu^\ell_B$ of the frequency shift can be computed from a field strength and stellar model (which enters Equation \ref{numag} purely through $\mathscr{I}$).

The weighted average $\langle B_r^2\rangle$ which appears in Equation \ref{numag} is given by \citet{li2022magnetic}:
\begin{equation} \label{avgfield}
    \langle B_r^2\rangle = \int_{\mathcal{R}^\ell_{\;\nu}}\mathrm{d}r\,K(r)\int \frac{\mathrm{d}\Omega}{4\pi}\,B_r^2\mathrm{,}
\end{equation}

\noindent where the weight function is

\begin{equation} \label{weightfunction}
    K(r) \simeq \begin{cases}
        \frac{N^3/\rho r^3}{\int_{\mathcal{R}^\ell_{\;\nu}}(N^3/\rho r^3)\,\mathrm{d}r} & \mathrm{in}\,\mathcal{R}^\ell_{\;\nu} \\
        0 & \mathrm{otherwise}
    \end{cases}
    \mathrm{.}
\end{equation}

The quantity $\nu^\ell_B$ is the mean frequency shift averaged over a multiplet.
However, individual modes within the multiplet will experience different frequency shifts from each other.
The precise order-unity prefactor relating a mode's magnetic frequency shift to $\nu^\ell_B$ depends both on the quantum numbers of the mode and the geometry of the field \citep{li2022magnetic}.
As a consequence, unlike rotation, magnetism does not generically preserve the symmetry of the multiplet, and the asymmetry of the multiplet can distinguish frequency shifts due to magnetism from those due to first-order rotation.
\new{We assume throughout this work that the pulsations are aligned with the rotation axis, so that each mode is observed as a single sinusoidal periodicity.
This is appropriate when the magnetic force is weaker than the Coriolis force (although see Section \ref{cowd}).}

The asymmetry of a multiplet can be parameterized by dimensionless asymmetry parameters $a^\ell_{m_1\,m_2\,m_3}$, the most commonly used of which (``$a$'') is defined as

\begin{equation} \label{dipoleasym}
    a^{\ell=1}_{^-\!1\,0\,^+\!1} = \frac{\Delta^{\ell=1}_{^-\!1\,0\,^+\!1}}{3\,\nu^{\ell=1}_B}\mathrm{,}
\end{equation}

\noindent where
\begin{equation} \label{dipolefreqcombo}
    \Delta^{\ell=1}_{^-\!1\,0\,^+\!1} = \nu^{\ell=1}_{m=^-\!1} -2\,\nu^{\ell=1}_{m=0} + \nu^{\ell=1}_{m=^+\!1}
\end{equation}

\noindent is a linear combination of the three dipole modes in a triplet \citep{li2022magnetic,mathis2023asymmetries,das2024unveiling}.
This is the only possible asymmetry parameter for $\ell=1$.

We similarly construct quadrupole asymmetry parameters by taking linear combinations $\Delta^\ell_{m_1\,m_2\,m_3}\propto\nu^\ell_B$ of three modes within the same multiplet such that the first-order rotational splitting terms cancel out (our adopted definitions for $\Delta^{\ell=2}_{m_1\,m_2\,m_3}$ are given by Equations \ref{allquadrupoles}).
We define the ten quadrupole asymmetry parameters $a^\ell_{m_1\,m_2\,m_3}$ as
\begin{equation} \label{quadrupoleasym}
    a^\ell_{m_1\,m_2\,m_3} = \frac{\Delta^{\ell=2}_{m_1\,m_2\,m_3}}{5\,\nu^{\ell=2}_B}\mathrm{.}
\end{equation}
In Appendix \ref{formalism}, we relate them to the magnetic field geometry and compute them for an inclined dipolar magnetic field \citep[see also][]{das2024unveiling}.

In WDs, $\rho$ sharply decreases towards surface, and $N$ often also reaches its peak value there, just below the surface convective zone.
The weight function $K(r)$ is thus very sharply peaked just below the surface, at depths $1-r/R \sim 10^{-3}$--$10^{-2}$, and asteroseismology is most sensitive to the near-surface field (\textit{center panel} of Figure \ref{fig:fancy_profile_main}).
This is in contrast to the cases of red giant or intermediate-mass main-sequence stars, within which $K(r)$ peaks at some highly stratified layer in the deep interior of the star \citep[the hydrogen-burning shell and near-core composition gradient, respectively;][]{fuller2015asteroseismology,lecoanet2022asteroseismic}.
While the pulsations of WDs are in principle modified by deep internal magnetic fields, the influence of these fields is highly diluted.
In the $0.64M_\odot$ WD model in Figure \ref{fig:fancy_profile_main}, a uniform magnetic field of $\simeq1\,\mathrm{MG}$ restricted to the inner $r/R\lesssim0.993$ of the WD affects a $P=1200\,\mathrm{s}$ gravity mode identically to a uniform $\simeq1\,\mathrm{G}$ field restricted to the outer layers with $r/R\gtrsim0.993$.
Long-period modes probe the magnetic field at very low-density layers ($\rho\simeq4\times10^{-6}\,\mathrm{g}\,\mathrm{cm}^{-3}$ for the $P=1200\,\mathrm{s}$ mode in Figure \ref{fig:fancy_profile_main}), although still far within the photosphere of the WD.

In principle, magnetic shifts also modify the spacing between adjacent multiplets, causing a period-dependent period spacing $\Delta\Pi_\ell=\Delta\Pi_\ell(P)$ \citep[e.g.,][]{cantiello2016asteroseismic,rui2024asteroseismic}.
\citet{deheuvels2023strong} used this effect to measure strong magnetic fields in red giants in which dipole triplets were not detected.
However, this is likely difficult in WDs, whose g modes already depart from period-uniformity due to non-asymptotic mode trapping effects due to, e.g., near-surface composition gradients \citep{brassard1992adiabatic}.
We instead focus on intra-multiplet asymmetries, since non-asymptotic effects are expected to affect each component of a multiplet in the same way.

\subsection{Mode suppression} \label{sectmodesuppression}

About $\simeq20\%$ of observed red giants have unusually low-amplitude non-radial (dipole) oscillations \citep{garcia2014study,stello2016prevalence}.
\citet{fuller2015asteroseismology} show that the amplitudes of these suppressed oscillations are consistent with nearly total damping of g-mode oscillations in the core.
They argue that such near-total suppression of g modes can occur if the damping mechanism is magnetic in nature.
Simple scaling arguments \citep{fuller2015asteroseismology,cantiello2016asteroseismic,rui2023gravity} show that g modes will be significantly affected by a magnetic field when the radial component of the field approaches $B_{r,\mathrm{crit}}$ as defined in Equation \ref{bcrit}, and detailed calculations validate several magnetic dissipation mechanisms \citep{lecoanet2017conversion,loi2017torsional,loi2018effects,loi2020effect,lecoanet2022asteroseismic,rui2023gravity}.
Throughout this work, we assume that magnetic g-mode suppression damps out all g-mode energy \citep[although see][]{mosser2017dipole}, and use the presence of non-suppressed g modes in WDs to set upper bounds on the magnetic field.

As can be seen in Equation \ref{bcrit}, longer-period modes require weaker magnetic fields to suppress.
At fixed $\ell$, the longest-period modes observed in a WD therefore set the strongest upper limits on the WD field.
Also, since $B_{r,\mathrm{crit}}\propto\sqrt{\rho}/N$, the value of $B_{r,\mathrm{crit}}$ reaches a minimum near the outer edge of the WD g-mode cavity.
Therefore, like magnetic asymmetries, magnetic mode suppression probes the near-surface layers.
The bottom panel of Figure \ref{fig:fancy_profile_main} shows $B_{r,\mathrm{crit}}$ for various g modes in a $0.64M_\odot$ WD.
While fields of $\simeq3\,\mathrm{G}$ at a depth of $r/R=0.998$ are sufficient to suppress $P=1200\,\mathrm{s}$ oscillations, a $\simeq10\,\mathrm{MG}$ field is needed to achieve the same effect at a depth of $r/R=0.9$.

For simplicity, when calculating $B_{r,\mathrm{crit}}$, we adopt the dimensionless prefactors shown in Equation \ref{bcrit} originally derived by \citet{fuller2015asteroseismology}.
In reality, this prefactor actually depends on $\ell$, $m$, and the field geometry in complicated ways \citep{lecoanet2017conversion,loi2018effects,rui2023gravity}.
Because this prefactor is order-unity \citep{lecoanet2017conversion,rui2023gravity}, its uncertain value translates to order-unity errors in the inferred field strength, but changes in its value are unlikely to greatly affect the field constraint.

\section{Methods} \label{methods}

\begin{figure*}
    \centering
    \includegraphics[width=\textwidth]{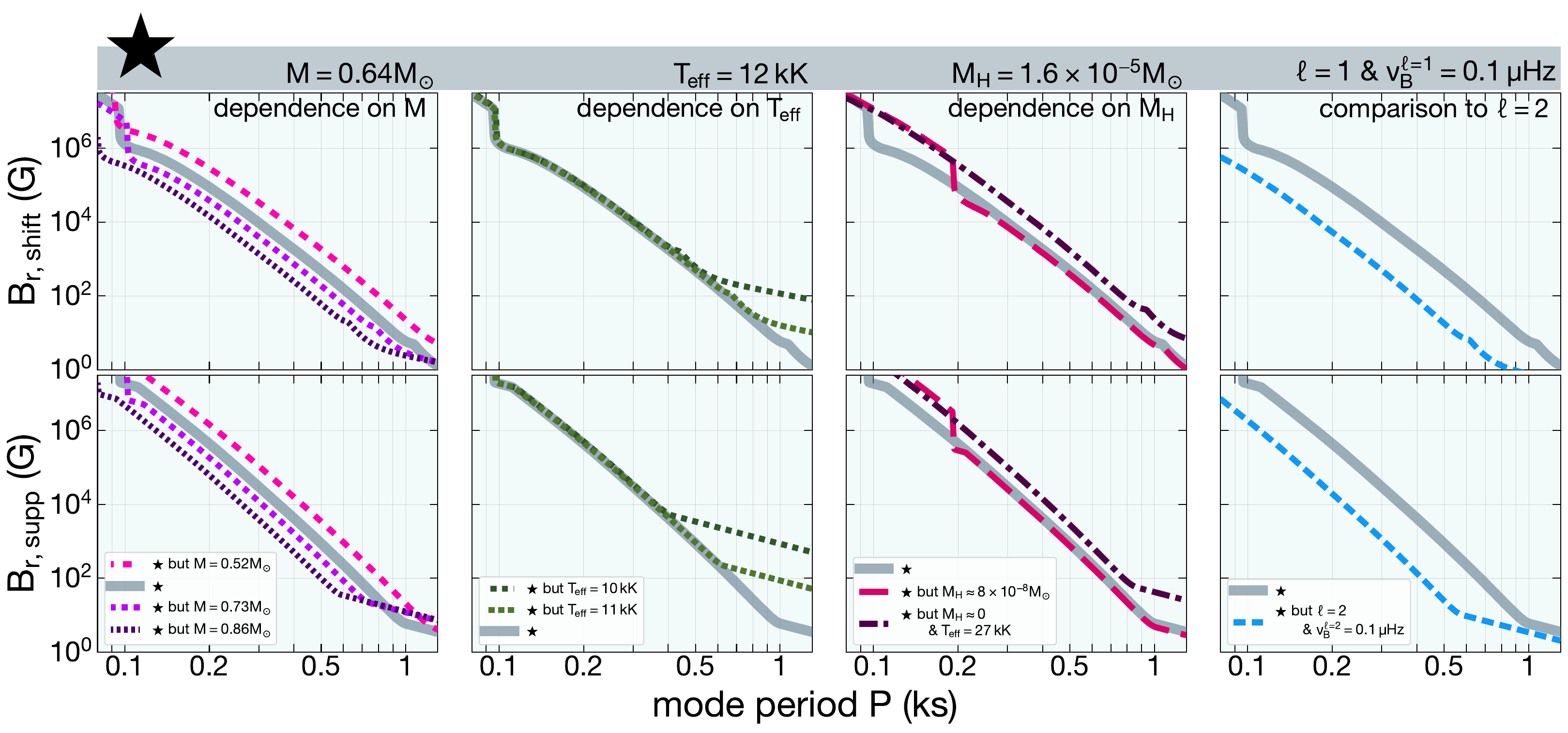}
    \caption{Model seismic magnetic field sensitivities $B_{r,\mathrm{shift}}$ and $B_{r,\mathrm{supp}}$.
    From \textit{left} to \textit{right}, columns show the effect of varying total mass $M$, effective temperature $T_{\mathrm{eff}}$, hydrogen atmosphere mass $M_{\mathrm{H}}$, and angular degree $\ell$.
    These comparisons are made relative to fiducial parameters (listed in the top \textit{gray} box), which we plot in \textit{gray} and denote with $\bigstar$.
    In particular, we show the value of $B_{r,\mathrm{shift}}$ implied by a dipole multiplet with an asymmetry of $\nu^{\ell=1}_B=0.1\,\mu\mathrm{Hz}$.
    Note that $B_{r,\mathrm{shift}}\propto\sqrt{\nu^\ell_B}$ whereas $B_{r,\mathrm{supp}}$ does not depend on $\nu^\ell_B$ at all.}
    \label{fig:show_constraining_power_polish}
\end{figure*}

\subsection{White dwarf models} \label{modelsect}

Seismic magnetic field estimates rely on detailed stellar models.
To this end, we use several helium-core and carbon--oxygen WD models prepackaged with version r10398 of the Modules for Experiments in Stellar Astrophysics code \citep[MESA;][]{paxton2010modules,paxton2013modules,paxton2015modules,paxton2018modules}.
We evolve models down a cooling track in the presence of gravitational settling using as diffusion representatives $^{1}\mathrm{H}$, $^{4}\mathrm{He}$, $^{12}\mathrm{C}$, $^{16}\mathrm{O}$, $^{20}\mathrm{Ne}$, and $^{22}\mathrm{Ne}$.
In order to avoid helium shell flashes and/or help with numerical convergence issues, we sometimes artificially reduce the size of the hydrogen atmosphere by replacing its inner layers with helium.
In one case ($M=0.73M_\odot$), we replace the outer layers of the originally helium-atmosphere model with hydrogen to create a DA WD model.
We also evolve several helium-atmosphere (DB) and DA WD models with lower total hydrogen masses $M_{\mathrm{H}}$, initialized by replacing some of the WD's hydrogen with helium prior to evolution.

We apply the Ledoux criterion for convective stability.
This sometimes produces short-lived convective zones owing to the particular composition gradient in the initial stellar model.
Although these convective zones can technically split the WD into multiple g mode cavities \citep[requiring more sophisticated analysis][]{pinccon2022multi}, the appearance times, locations, and lifetimes of these convective zones are hard to predict.
When performing integrals over the g mode cavity (i.e., when determining $\mathcal{R}^\ell_{\;\nu}$), we therefore integrate over all shells satisfying the gravity wave propagation condition ($2\pi\nu<N,S_\ell$) without enforcing that the g-mode cavity is contiguous.
Our WD models are summarized in Table \ref{tab:wd_models}.\footnote{These MESA models can be reproduced with the inlists, \texttt{run\_star\_extras.f}, and other files at the following link: \new{\dataset[https://doi.org/10.5281/zenodo.14457127]{https://doi.org/10.5281/zenodo.14457127}}}

\begin{table}[]
    \centering
    \begin{tabular}{llllll}
        \hline
        core & type & $M$ ($M_\odot$) & $M_{\mathrm{H}}$ ($M_\odot$) & $M_{\mathrm{He}}$ ($M_\odot$) \\
        \hline
        He & DA & $0.15^*$ & $9.95\times10^{-4}$ & \multicolumn{1}{c}{$\cdots$} \\
        He & DA & $0.20$ & $9.54\times10^{-4}$ & \multicolumn{1}{c}{$\cdots$} \\
        He & DA & $0.40^*$ & $7.26\times10^{-5}$ & \multicolumn{1}{c}{$\cdots$} \\
        CO & DA & $0.52$ & $8.53\times10^{-5}$ & $2.90\times10^{-2}$ \\
        CO & DB & $0.52$ & $0$ & $2.91\times10^{-2}$ \\
        CO & DA & $0.64$ & $1.64\times10^{-5}$ & $4.23\times10^{-3}$ \\
        CO & DA & $0.64^*$ & $8.00\times10^{-8}$ & $4.25\times10^{-3}$ \\
        CO & DB & $0.64$ & $0$ & $4.25\times10^{-3}$ \\
        CO & DA & $0.73$ & $3.48\times10^{-6}$ & $2.53\times10^{-3}$ \\
        CO & DB & $0.73$ & $0$ & $2.54\times10^{-3}$ \\
        CO & DA & $0.86$ & $3.99\times10^{-6}$ & $1.64\times10^{-3}$ \\
        CO & DB & $0.86$ & $0$ & $1.65\times10^{-2}$ \\
        \hline
    \end{tabular}
    \caption{Summary of the WD models used in this work (Section \ref{modelsect}).
    Asterisks ($^*$) indicate that we checked the magnetic sensitivity of the model, but did not use it in fitting WD observations.}
    \label{tab:wd_models}
\end{table}

\subsection{Seismic field constraints} \label{seismicfieldconstraints}

Magnetism tends to asymmetrize rotational multiplets.
Symmetric multiplets therefore place upper limits on the magnetic field present in a WD.
For each multiplet, we use Equations \ref{dipoleasym} (for $\ell=1$) or \ref{quadrupoleasym} (for $\ell=2$) to convert a measured asymmetry to $\nu^\ell_B$.
To be conservative, we calculate this magnetic field bound assuming that the measured asymmetry is underestimated by $2\sigma$, i.e., using as the asymmetry $\lvert\Delta^\ell_{m_1\,m_2\,m_3}\rvert+2\delta\Delta^\ell_{m_1\,m_2\,m_3}$ where $\delta\Delta^\ell_{m_1\,m_2\,m_3}$ is the uncertainty on the asymmetry implied by the measurement uncertainties on the mode frequencies.
Finally, we invert Equation \ref{numag} to solve for the upper limit $B_{r,\mathrm{shift}}=\sqrt{\langle B_r^2\rangle}$:
\begin{equation} \label{shiftshift}
    B_{r,\mathrm{shift}} = 8\pi^{5/2}\sqrt{\mathrm{min}\left\lbrace\frac{\lvert\Delta^\ell_{m_1\,m_2\,m_3}\rvert+2\delta\Delta^\ell_{m_1\,m_2\,m_3}}{\lvert a^\ell_{m_1\,m_2\,m_3}\rvert A_\ell\mathscr{I}P^3}\right\rbrace}
\end{equation}

\noindent where the minimum is over all combinations of three modes (indexed by $m_1$, $m_2$, and $m_3$) within the same multiplet.
The quantity $\mathscr{I}$ is corrected for non-asymptotic effects using an ad hoc factor described in Appendix \ref{sectnonasymptotic}.

In this work, we assume asymmetry parameters $a^\ell_{m_1\,m_2\,m_3}$ typical for an inclined, centered dipolar magnetic field (derived in Appendix \ref{inclined}).
This sets $a^{\ell=1}_{^-\!1\,0\,^+\!1}\approx0.253$ and $a^{\ell=2}_{m_1\,m_2\,m_3}$ to values between $-0.217$ and $-0.036$, depending on the values of $m_1$, $m_2$, and $m_3$.
While the field geometry is uncertain in reality, changes to the field geometry only modify the field strength estimate by order-unity factors.
Mode asymmetries can also be non-magnetic in origin (e.g., from second-order rotational effects or near-degeneracy effects).
However, we assume that other sources of asymmetry do not decrease the rough scale of the observed symmetry.
In principle, we can underestimate the magnetic field if another source of asymmetry happens to nearly cancel out the magnetic asymmetry.

When $B_r>B_{r,\mathrm{crit}}(P)$, a g mode of period $P$ is expected to be suppressed.
Therefore, the presence of a g mode indicates that the field does not exceed $B_{r,\mathrm{crit}}$.
We use this fact to place additional suppression-based upper bounds $B_{r,\mathrm{supp}}$ by setting $B_{r,\mathrm{supp}}=B_{r,\mathrm{crit}}$ and inverting Equation \ref{bcrit} for each mode:
\begin{equation} \label{suppsupp}
    B_{r,\mathrm{supp}} = 4\pi^{5/2}\mathrm{min}\left\lbrace\frac{\left(\sqrt{\rho}\,r/N\right)_{\mathrm{min}\,g,{P_0}}}{\sqrt{\ell(\ell+1)}P_0^2}\right\rbrace
\end{equation}

\noindent where the minimum is over all observed radial orders.
The quantity $\left(\sqrt{\rho}\,r/N\right)_{\mathrm{min}\,\mathrm{g},P}$ is the minimum value of $\sqrt{\rho}\,r/N$ in the g-mode cavity corresponding to period $P$.
The period $P_0$ indicates the central frequency of the multiplet, which we take as an estimate for the unperturbed mode period.
When the $m=0$ component is reported, we take its period to be $P_0$.
Otherwise, if two or more components are reported, we calculate $P_0=1/\nu^{(0)}$, where $\nu^{(0)}$ is the $y$-intercept of a fit to the mode frequencies $\nu$ within a multiplet to a linear function of $m$.
Finally, if there is only a single mode within the multiplet, we take its period as $P_0$.
As can be seen in Equation \ref{bcrit}, the minimum is always set by either the longest-period dipole multiplet or longest-period quadrupole multiplet.
Since this constraint relies on the \textit{existence} of an excited radial order rather than a rotational multiplet asymmetry, we only require a single observed mode within a multiplet.

\subsection{Dependence of magnetic sensitivity on white dwarf properties}

The sensitivity of a WD's oscillation modes to magnetism depends on the WD's structure, particularly through $\mathscr{I}$ (defined in Equation \ref{curlyIasympt}) for $B_{r,\mathrm{shift}}$ and $(\sqrt{\rho}\,r/N)_{\mathrm{min}\,g,P}$ (through Equation \ref{suppsupp}) for $B_{r,\mathrm{supp}}$.

The first three columns of Figure \ref{fig:show_constraining_power_polish} show how $B_{r,\mathrm{shift}}$ and $B_{r,\mathrm{supp}}$ depend on various WD properties relative to a fiducial DAV model with $M=0.64M_\odot$ and $T_{\mathrm{eff}}=12\,\mathrm{kK}$.
To good approximation, both $B_{r,\mathrm{shift}}$ (at fixed $\nu^\ell_B$) and $B_{r,\mathrm{supp}}$ can be described as broken power laws, with a shallower power law index at higher mode periods.
The transition period $P_t$ between these two regimes corresponds to the mode period above (below) which the outer boundary g-mode cavity is set by $N$ ($S_\ell$).
It is given by $P_t=2\pi/S_\ell$ at the outer intersection between $N$ and $S_\ell$, and corresponds to a long period $P_t\simeq1100\,\mathrm{s}$ ($\nu\simeq900\,\mu\mathrm{Hz}$) for dipole modes in the model in Figure \ref{fig:fancy_profile_main} \citep[consistent with the value computed by][]{montgomery2020limits}.
Hereafter, we refer to modes with long periods $P>P_t$ (short periods $P<P_t$) as $N$-limited ($S_\ell$-limited).

The Lamb frequency $S_\ell$ has a gradually varying profile.
This means the outer boundary of the g-mode cavity varies substantially with $P$ for $S_\ell$-limited modes.
Since the layers of the WD most sensitive to magnetism are those closest to the outer boundary of the radiative zone, both $B_{r,\mathrm{shift}}$ and $B_{r,\mathrm{supp}}$ for $S_\ell$-limited modes strongly depend on $P$.
Conversely, the Brunt--V\"ais\"al\"a frequency $N$ drops sharply at the outer boundary of the radiative zone.
Accordingly, the outer g-mode-cavity boundaries of $N$-limited modes are essentially identical to each other.
In this regime, $B_{r,\mathrm{shift}}$ and $B_{r,\mathrm{supp}}$ have the approximate scalings $B_{r,\mathrm{shift}}\propto P^{-3/2}$ and $B_{r,\mathrm{supp}}\propto P^{-2}$, reflecting only the explicit period dependences in Equations \ref{shiftshift} and \ref{suppsupp}.

More massive WDs have higher dynamical frequencies, which in turn increases both $N$ and $S_\ell$ (\textit{first column} of Figure \ref{fig:show_constraining_power_polish}).
In the $S_\ell$-limited regime, the increased $S_\ell$ in more massive WDs translates to an outer g-mode cavity boundary which is closer to the surface.
$S_\ell$-limited modes in more massive WDs are thus more magnetically sensitive than those in less massive WDs.
We find that the magnetic sensitivity of $N$-limited modes, however, is almost independent of mass.

The primary effect of varying the effective temperature $T_{\mathrm{eff}}$ (\textit{second column} of Figure \ref{fig:show_constraining_power_polish}) is to change the extent of the outer, partial ionization-driven convection zone.
As DA WDs cool below $T_{\mathrm{eff}}\sim13\,\mathrm{kK}$, they develop progressively deeper (albeit low mass; $M_{\mathrm{conv}}/M_{\mathrm{WD}}\sim10^{-12}$) convective zones driven by a near-surface partial ionization zone \citep[e.g.,][]{tremblay2015calibration}.
The result is that both $B_{r,\mathrm{shift}}$ and $B_{r,\mathrm{supp}}$ are unaffected for $S_\ell$-limited modes.
However, $N$-limited modes are less sensitive to magnetic fields in cooler WDs because they are confined deeper in the WD where $\rho$ is larger and $N$ is smaller.
Furthermore, cooler WDs have lower $P_t$: as they cool, the range of frequencies which are $N$-limited widens.
While only very-long dipole period modes are $N$-limited in most cases, $P_t$ drops dramatically near the very red edge of the DAV instability strip ($P_t\simeq600\,\mathrm{s}$ in the $T_{\mathrm{eff}}=10\,\mathrm{kK}$ model in the \textit{second panel} of Figure \ref{fig:show_constraining_power_polish}).

Because the near-surface partial ionization zone is also responsible for mode excitation \citep{brickhill1991pulsations,goldreich1999gravity}, the DA instability strip contains DAVs with both deep and shallow convective zones on its red and blue sides, respectively.
Deep convective zones also develop within the the DBV instability strip for similar reasons, albeit at a hotter temperature $T_{\mathrm{eff}}\approx30\,\mathrm{kK}$ \citep{fontaine1987recent}.
Diminished sensitivity of long-period modes in cooler WDs (relative to hotter ones) trades off with the tendency of cooler WDs to excite more sensitive longer-period modes \citep[e.g.,][]{van2012newly,hermes2017white}.

Although hydrogen spectral lines in DA WDs indicate the presence of surface hydrogen, the precise amount of this hydrogen is very uncertain.
The \textit{third column} of Figure \ref{fig:show_constraining_power_polish} shows that the effect of changing the hydrogen mass $M_{\mathrm{H}}$ is minor.
This is because the depth of the surface convective zone is largely insensitive to $M_{\mathrm{H}}$, as long as it is enough to prevent the development of a helium partial ionization-driven convection zone around $\simeq25\,\mathrm{kK}$.
Figure \ref{fig:show_constraining_power_polish} also shows the sensitivity of a DB model in the DBV instability strip which, despite being much hotter, has a similar sensitivity to the DAV model.

At fixed $\nu^\ell_B$, quadrupole modes are more sensitive than dipole modes (\textit{last column} in Figure \ref{fig:show_constraining_power_polish}), especially for $S_\ell$-limited modes.
Since $S_\ell\propto\sqrt{\ell(\ell+1)}$, the outer boundary of the g-mode cavity of $S_\ell$-limited quadrupole modes is farther out than those of $S_\ell$-limited dipole modes.
For the same reason, $P_t$ is significantly lower for quadrupole modes ($P_t\simeq600\,\mathrm{s}$ for the model in Figure \ref{fig:show_constraining_power_polish}


In contrast, $N$-limited quadrupole modes only achieve a modest improvement in sensitivity over their dipole counterparts.
In this case, both $B_{r,\mathrm{shift}}$ and $B_{r,\mathrm{supp}}$ are decreased by a factor of $\approx\sqrt{3}$, owing to their explicit dependences on $\ell$ in Equations \ref{shiftshift} and \ref{suppsupp}.

\section{Seismic field constraints in observed white dwarfs} \label{seismicconstraints}

\begin{figure}
    \centering
    \includegraphics[width=0.46\textwidth]{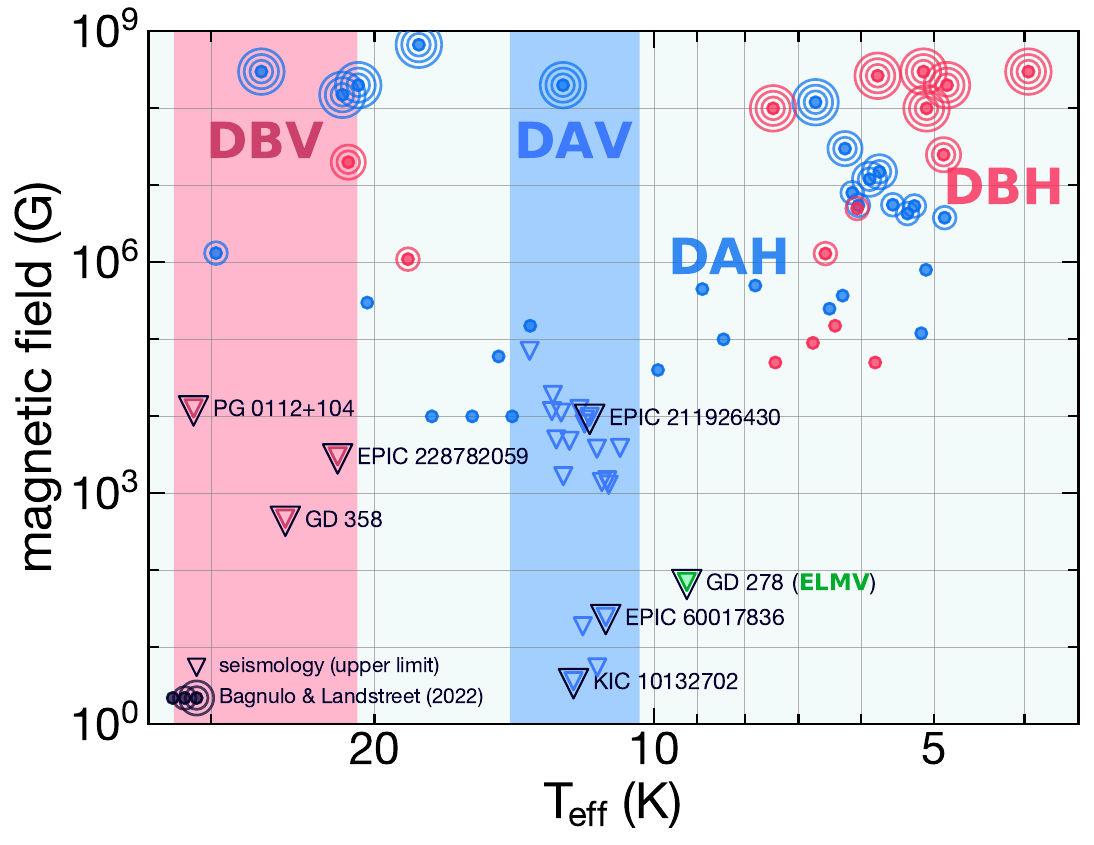}
    \caption{Seismic upper limits on WD magnetic fields $\min(B_{r,\mathrm{shift}},B_{r,\mathrm{supp}})$ (\textit{triangles}) versus $T_{\mathrm{eff}}$, compared to the spectropolarimetric measurements (\textit{circles}) of \citet{bagnulo2022multiple}.
    \textit{Concentric circles} have the same meaning as in Figure \ref{fig:wd_summary}.
    We explicitly label WDs whose leaf diagrams are shown in Figures \ref{fig:fieldplot_DAV}, \ref{fig:fieldplot_DBV}, and \ref{fig:fieldplot_ELMV}.}
    \label{fig:total_field_summary}
\end{figure}

We place asymmetry- and suppression-based upper bounds on the magnetic field for 24 observed pulsating WDs of varying types (Table \ref{tab:fieldconstraints}).
For each WD, we first identify the stellar model which best minimizes
\begin{equation}
    \chi^2 = \left(\frac{T_{\mathrm{eff,spec}}-T_{\mathrm{eff,model}}}{100\,\mathrm{K}}\right)^2 + \left(\frac{\log g_{\mathrm{spec}} - \log g_{\mathrm{model}}}{0.01}\right)^2\mathrm{,}
\end{equation}

\noindent where $T_{\mathrm{eff,spec}}$ and $\log g_{\mathrm{eff,spec}}$ are spectroscopically measured values of $T_{\mathrm{eff}}$ and $\log g$ taken from the literature.
In other words, we choose the stellar model which best matches the location of the WD on the Kiel diagram.
Using this model, we then compute $B_{r,\mathrm{shift}}$ and $B_{r,\mathrm{supp}}$ using Equations \ref{shiftshift} and \ref{suppsupp}.
These upper bounds are shown in Table \ref{tab:fieldconstraints} and Figure \ref{fig:total_field_summary}.

\begin{table*}
\caption{Observed and model spectroscopic/seismic properties and upper limits on the magnetic field of $24$ WDs}\label{tab:fieldconstraints}

\begin{tabular}{l|ll|ll|ll|ll|ll}
\toprule
name & type & model $M$ & \multicolumn{2}{c}{$P$ (ks)} & \multicolumn{2}{c}{$T_{\mathrm{eff}}$ (kK)} & \multicolumn{2}{c}{$\log g$} & \multicolumn{2}{c}{$B_r$ upper bound (G)} \\
\cmidrule(lr){1-1}\cmidrule(lr){2-3}\cmidrule(lr){4-5}\cmidrule(lr){6-7}\cmidrule(lr){8-9}\cmidrule(lr){10-11}
& & & $P_{\mathrm{min}}$ & $P_{\mathrm{max}}$ & data & model & data & model & $B_{r,\mathrm{shift}}$ & $B_{r,\mathrm{supp}}$ \\
\midrule
GD 278\textsuperscript{1,2} & ELMV & 0.20 & $2.29$ & $6.73$ & $9.23$ & $9.33$ & $6.63$ & $6.65$ & $100$ & \new{$\underline{70}$} \\
KIC 4357037\textsuperscript{3} & DAV & 0.64 & $0.28$ & $0.35$ & $12.75$ & $12.81$ & $8.02$ & $8.07$ & \new{$\underline{4\mathrm{k}}$} & $10\mathrm{k}$ \\
KIC 4552982\textsuperscript{3} & DAV & 0.73 & $0.36$ & $0.36$ & $11.24$ & $11.15$ & $8.28$ & $8.24$ & \new{$\underline{1\mathrm{k}}$} & $4\mathrm{k}$ \\
KIC 7594781\textsuperscript{3} & DAV & 0.64 & $0.28$ & $0.30$ & $12.04$ & $11.93$ & $8.17$ & $8.07$ & $\underline{10\mathrm{k}}$ & $30\mathrm{k}$ \\
KIC 10132702\textsuperscript{3} & DAV & 0.73 & $0.62$ & $0.89$ & $12.22$ & $12.20$ & $8.17$ & $8.23$ & $8$ & \new{$\underline{4}$} \\
KIC 11911480\textsuperscript{3} & DAV & 0.64 & $0.20$ & $0.32$ & $11.88$ & $11.93$ & $8.02$ & $8.07$ & \new{$\underline{8\mathrm{k}}$} & $20\mathrm{k}$ \\
EPIC 60017836\textsuperscript{3} & DAV & 0.73 & $0.83$ & $1.22$ & $11.28$ & $11.34$ & $8.14$ & $8.23$ & $50$ & \new{$\underline{20}$} \\
EPIC 201719578\textsuperscript{3} & DAV & 0.64 & $0.37$ & $0.46$ & $11.39$ & $11.39$ & $8.07$ & $8.08$ & \new{$\underline{1\mathrm{k}}$} & $2\mathrm{k}$ \\
EPIC 201730811\textsuperscript{3} & DAV & 0.64 & $0.16$ & $0.35$ & $12.60$ & $12.63$ & $7.96$ & $8.07$ & \new{$\underline{10\mathrm{k}}$} & \new{$\underline{10\mathrm{k}}$} \\
EPIC 201802933\textsuperscript{3} & DAV & 0.73 & $0.12$ & $0.40$ & $12.53$ & $12.51$ & $8.14$ & $8.23$ & \new{$\underline{2\mathrm{k}}$} & \new{$\underline{2\mathrm{k}}$} \\
EPIC 201806008\textsuperscript{3} & DAV & 0.73 & $0.41$ & $0.41$ & $11.20$ & $11.15$ & $8.18$ & $8.24$ & \new{$\underline{1\mathrm{k}}$} & \new{$\underline{1\mathrm{k}}$} \\
EPIC 210397465\textsuperscript{3} & DAV & 0.52 & $0.67$ & $1.39$ & $11.52$ & $11.56$ & $7.78$ & $7.85$ & $200$ & \new{$\underline{5}$} \\
EPIC 211596649\textsuperscript{3} & DAV & 0.64 & $0.27$ & $0.30$ & $11.89$ & $11.93$ & $7.97$ & $8.07$ & \new{$\underline{7\mathrm{k}}$} & $30\mathrm{k}$ \\
EPIC 211629697\textsuperscript{3} & DAV & 0.52 & $0.49$ & $0.49$ & $10.89$ & $10.85$ & $7.95$ & $7.85$ & \new{$\underline{3\mathrm{k}}$} & $4\mathrm{k}$ \\
EPIC 211914185\textsuperscript{3} & DAV & 0.86 & $0.11$ & $0.20$ & $13.62$ & $13.66$ & $8.44$ & $8.42$ & \new{$\underline{60\mathrm{k}}$} & $80\mathrm{k}$ \\
EPIC 211926430\textsuperscript{3} & DAV & 0.64 & $0.12$ & $0.30$ & $11.74$ & $11.56$ & $8.06$ & $8.08$ & \new{$\underline{9\mathrm{k}}$} & $30\mathrm{k}$ \\
EPIC 228682478\textsuperscript{3} & DAV & 0.73 & $0.29$ & $0.29$ & $12.34$ & $12.20$ & $8.23$ & $8.23$ & \new{$\underline{4\mathrm{k}}$} & $10\mathrm{k}$ \\
EPIC 229227292\textsuperscript{3} & DAV & 0.64 & $0.29$ & $0.37$ & $11.53$ & $11.56$ & $8.15$ & $8.08$ & \new{$\underline{3\mathrm{k}}$} & $7\mathrm{k}$ \\
EPIC 220204626\textsuperscript{3} & DAV & 0.73 & $0.51$ & $0.80$ & $11.94$ & $11.82$ & $8.26$ & $8.23$ & $60$ & \new{$\underline{20}$} \\
EPIC 220258806\textsuperscript{3} & DAV & 0.64 & $0.15$ & $0.30$ & $12.89$ & $12.81$ & $8.09$ & $8.07$ & \new{$\underline{10\mathrm{k}}$} & $30\mathrm{k}$ \\
EPIC 220347759\textsuperscript{3} & DAV & 0.64 & $0.12$ & $0.25$ & $12.86$ & $12.81$ & $8.09$ & $8.07$ & \new{$\underline{20\mathrm{k}}$} & $90\mathrm{k}$ \\
EPIC 228782059\textsuperscript{4} & DBV & 0.73 & $0.23$ & $0.34$ & $21.91$ & $21.76$ & $8.14$ & $8.23$ & $6\mathrm{k}$ & \new{$\underline{3\mathrm{k}}$} \\
GD 358\textsuperscript{5,6} & DBV & 0.52 & $0.42$ & $0.77$ & $24.94$ & $24.86$ & $7.75$ & $7.83$ & $700$ & \new{$\underline{500}$} \\
PG 0112+104\textsuperscript{7,8} & DBV & 0.52 & $0.15$ & $0.50$ & $31.30$ & $31.30$ & $7.80$ & $7.80$ & $200\mathrm{k}$ & \new{$\underline{10\mathrm{k}}$} \\
\bottomrule
\end{tabular}

For the $24$ WDs for which we place a magnetic field upper limit, the WD type, model mass $M$, minimum $P_{\mathrm{min}}$ and maximum $P_{\mathrm{max}}$ retained pulsation period, spectroscopically measured and model $T_{\mathrm{eff}}$ and $\log g$, and asymmetry- and suppression-based magnetic field upper limits $B_{r,\mathrm{shift}}$ and $B_{r,\mathrm{supp}}$ \new{(with the stronger upper limit underlined)}.
1: \citet{lopez2021discovery} 2: \citet{calcaferro2023exploring} 3: \citet{hermes2017white} 4: \citet{duan2021epic} 5: \citet{winget1994whole} 6: \citet{corsico2019pulsating} 7: \citet{hermes2017deep} 8: \citet{dufour2010multiwavelength}
\end{table*}

\subsection{Magnetic leaf diagrams}

\begin{figure*}
    \centering
    \includegraphics[width=\textwidth]{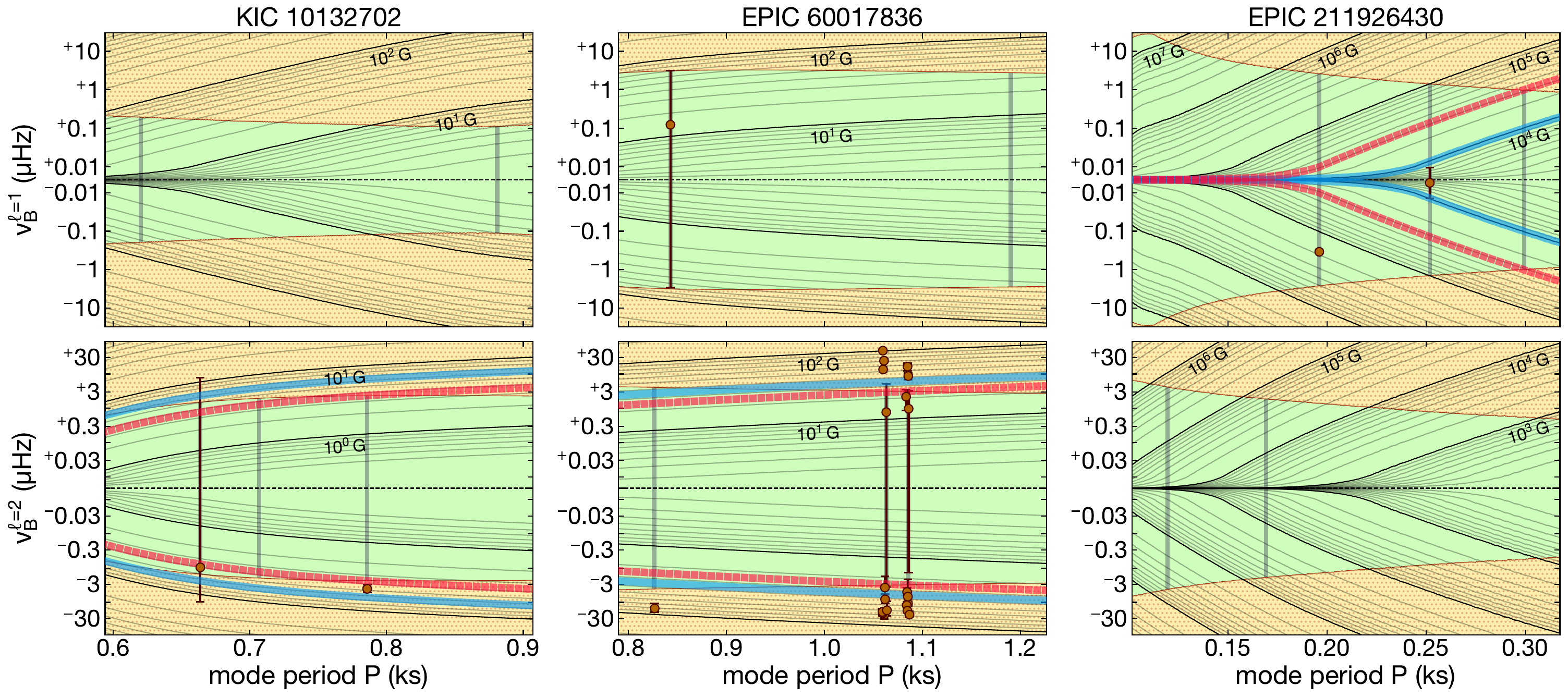}
    \caption{Leaf diagrams of three DAV WDs from \citet{hermes2017deep}.
    \textit{Brown circles} show measured asymmetry values, with error bars enclosing the $2\sigma$ confidence interval set by frequency measurement uncertainties.
    \textit{Gray vertical lines} indicate an estimate for the central frequency of a multiplet.
    \textit{Contours} indicate the degree of asymmetry implied by a given magnetic field.
    The \textit{yellow dotted region} indicates the forbidden region within which magnetic mode suppression is expected.
    The \textit{blue solid} contour indicates $B_{r,\mathrm{shift}}$: this is the lowest-field strength contour that fully encloses a single \textit{brown circle}, including its error bar.
    The \textit{red dashed} contour indicates $B_{r,\mathrm{supp}}$: this denotes the lowest-field strength contour that crosses all \textit{gray vertical lines} outside of the \textit{yellow dotted region}.}
    \label{fig:fieldplot_DAV}
\end{figure*}

We visualize the impact of magnetism on seismology using ``leaf diagrams,'' such as those shown in Figure \ref{fig:fieldplot_DAV}.
For each $\ell$, the leaf diagram plots as \textit{brown circles} the value of $\nu^\ell_B$ implied by a measurement of the asymmetry $\Delta^\ell_{m_1\,m_2\,m_3}$ (Equations \ref{dipoleasym} and \ref{quadrupoleasym}) against the mode period $P$.
The error bars on these points denote $2\sigma$ uncertainty intervals on the asymmetries.
The magnetic field required to reproduce an observed asymmetry can be read off of the diagram by comparing to the contours in the background of the leaf diagram, which show model predictions for $\nu^\ell_B$ (from Equation \ref{numag}).
The \textit{solid blue stripe} shows $B_{r,\mathrm{shift}}$, corresponding to the weakest-field strength contour which entirely encloses the lowest asymmetry point (the one closest to the $x$ axis), including its full error bar.

The \textit{yellow dotted regions} show ``magnetically forbidden'' values of $\nu^\ell_B$.
Observed asymmetries in this region are so high that the field required to generate them (in the perturbative theory) would also magnetically suppress the mode.
Under the present assumptions, such asymmetries must therefore be non-magnetic in origin.
Accordingly, if a field contour is in the forbidden region at some period $P$, that magnetic field strength suppresses a g mode of period $P$.
The boundary of the forbidden region is demarcated by $\nu^\ell_B$ evaluated at $B_{r,\mathrm{crit}}$.
The \textit{vertical gray lines} show the estimated central components of the multiplets, including those too incomplete for an asymmetry measurement.
The suppression constraint $B_{r,\mathrm{supp}}$ (the \textit{dotted red stripe}) can be read off the diagram by finding the weakest-field $\nu^\ell_B$ contour which intersects with all \textit{vertical gray lines} before entering the forbidden region.

As defined in Equation \ref{numag}, $\nu^\ell_B$ is strictly non-negative.
Negative values of $\nu^\ell_B$ on the leaf diagram represent those where the sign of the measured asymmetry differs from the assumed sign of $a^\ell_{m_1\,m_2\,m_3}$ (given for this work in Appendix \ref{inclined}).
\new{For example, \citet{li2022magnetic} show that $a^{\ell=1}_{^-\!1\,0\,^+\!1}$ can take a range of values from $-1/2$ to $1$, depending on the horizontal structure of the magnetic field.}
A negative value thus indicates an incorrect assumption about the magnetic field geometry, or a non-magnetic origin for the asymmetry altogether.
However, as $a^\ell_{m_1\,m_2\,m_3}$ is an order-unity constant, incorrect assumptions about its value only affect $B_{r,\mathrm{shift}}$ by order-unity factors.

\subsection{Carbon--oxygen white dwarfs} \label{cowd}

\begin{figure*}
    \centering
    \includegraphics[width=\textwidth]{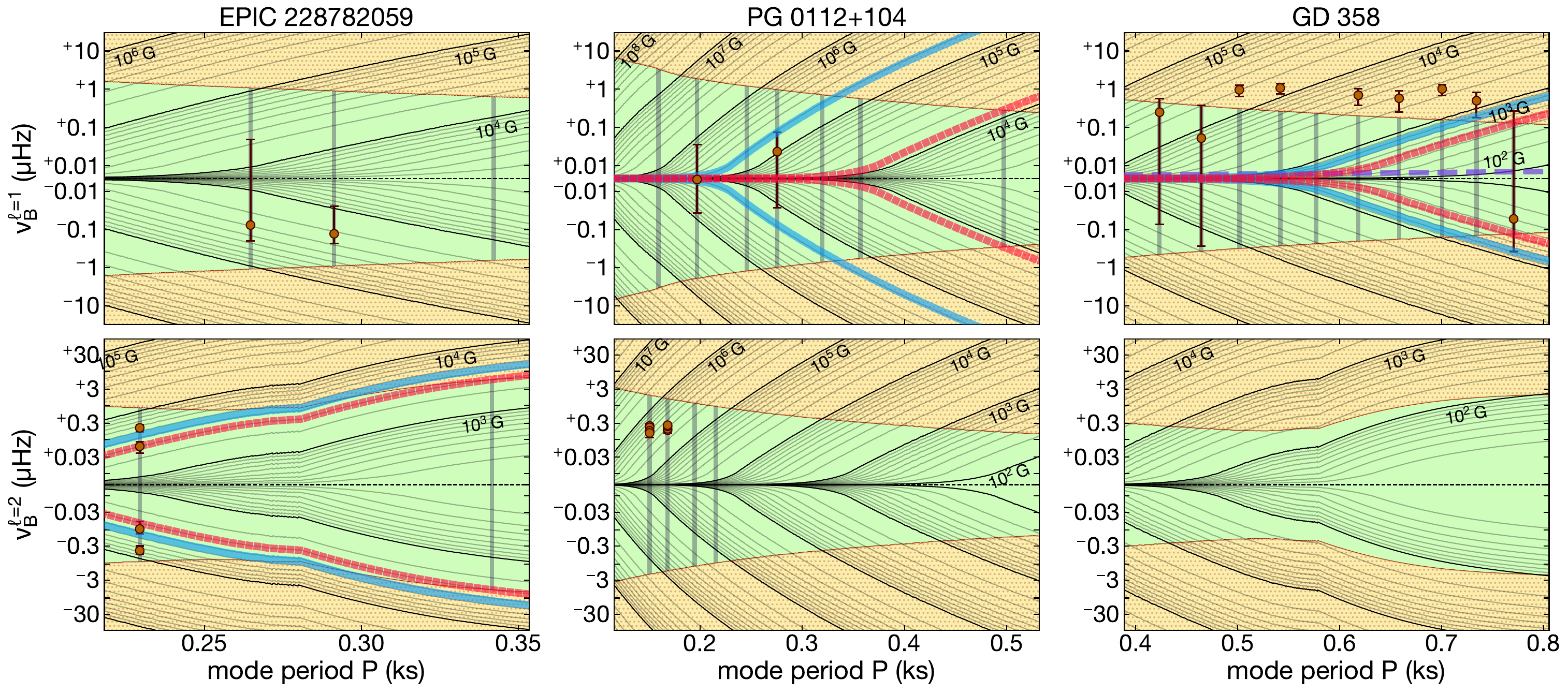}
    \caption{Same as Figure \ref{fig:fieldplot_DBV}, but for the DBV WDs EPIC 228782059 \citep{duan2021epic}, PG 0112+104 \citep{hermes2017deep}, and GD 358 \citep{winget1994whole}.
    The \textit{dashed purple curve} shows the predicted dipole asymmetry from second-order rotational effects.}
    \label{fig:fieldplot_DBV}
\end{figure*}

As Figure \ref{fig:wd_summary} shows, both the DAV and DBV WD instability strips lie blueward of both the concentration of magnetic WDs $\sim6\,\mathrm{kK}$ and the onset of carbon--oxygen crystallization for near-canonical masses $\simeq0.6M_\odot$.
Pulsating WDs may thus represent present-day magnetic WDs when they were younger and hotter.
It takes $>2\,\mathrm{Gyr}$ for both our $0.64M_\odot$ DA and DB models to cool from the DAV and DBV instability strips to $6\,\mathrm{kK}$.

Our sample of DAV WDs is taken from the \textit{K2} catalog of \citet{hermes2017white}.
Whenever possible, we use mode periods and uncertainties given by their nonlinear least-squares method.
However, they also find that some (usually low-period; $P\gtrsim800\,\mathrm{s}$) modes have power spectral linewidths which are significantly broader than implied by the spectral window, consistent with a lack of phase stability \citep[see also][]{winget1994whole,bischoff2019gd358,montgomery2020limits}.
In these cases, they fit a Lorentzian to the power spectral peak.
When the nonlinear least-squares uncertainties are not reported, we adopt the uncertainties on the mode period derived from this Lorentzian fit.

To be conservative, we have excluded uncertain or unknown identifications of $\ell$ for calculating $B_{r,\mathrm{supp}}$, as well as either $\ell$ or $m$ for $B_{r,\mathrm{shift}}$.
We have also excluded all modes which are labeled as combination frequencies.
This leaves 20 DAVs for which we place upper bounds on the magnetic field.
These upper bounds are usually in the range $1$--$10\,\mathrm{kG}$, but sometimes are as low as a few gauss for DAVs within which long-period modes are excited.
We show leaf diagrams for three selected DAVs in Figure \ref{fig:fieldplot_DAV}.

Additionally, we select $3$ DBVs which span the temperature range of the DBV instability strip.
Using mode frequencies reported by \citet{duan2021epic}, we place an upper bound $B_{r,\mathrm{shift}}\sim3\,\mathrm{kG}$ for the ``cool'' ($T_{\mathrm{eff}}\approx22\,\mathrm{kK}$) WD EPIC 228782059.
We similarly place an upper bound $B_{r,\mathrm{shift}}\sim10\,\mathrm{kG}$ for the ``hot'' \citep[$T_{\mathrm{eff}}\approx31\,\mathrm{kK}$;][]{dufour2010multiwavelength} DBV PG 0112+104, consistent with the bound placed by \citet{hermes2017deep}.
The leaf diagrams of these WDs are shown in Figure \ref{fig:fieldplot_DBV}.

As a further consistency check, we revisit GD 358, finding an asymmetry-based bound $B_{r,\mathrm{shift}}\sim800\,\mathrm{G}$ set by the lowest-frequency triplet (similar to their inferred value $\approx1.3\mathrm{kG}$), as well as a formally slightly stronger suppression-based bound $B_{r,\mathrm{supp}}\sim500\,\mathrm{G}$.
Curiously, subsequent observations of GD 358 have observed time variation in both the amplitudes and \textit{frequencies} of the oscillation modes \citep[e.g.,][]{bischoff2019gd358}.
The wandering of the frequencies takes place on timescales much longer than an observing season \citep[roughly several months;][]{provencal20092006}, and the frequencies vary by much more than the measured asymmetries of $\sim\!1 \, \mu$Hz shown in Figure \ref{fig:fieldplot_DBV}.
Both time-varying dynamo magnetic fields \citep{markiel1994dynamo} and convective zone thicknesses \citep{montgomery2010evidence,montgomery2020limits} have been proposed for this phenomenon, although a definitive explanation remains elusive.

Curiously, most of GD 358's asymmetries in Figure \ref{fig:fieldplot_DBV} lie in the magnetically forbidden region.
The degree of asymmetry is too large to be explained by second-order rotational effects \citep{dziembowski1992effects}, which predict $\Delta^{\ell=1}_{^-\!1\,0\,^+\!1}=P/20P_{\mathrm{rot}}^2\lesssim10^{-2}\,\mu\mathrm{Hz}$ (the \textit{dashed purple curve} in Figure \ref{fig:fieldplot_DAV}), much lower than the observed values $\Delta^{\ell=1}_{^-\!1\,0\,^+\!1}\sim0.5\,\mu\mathrm{Hz}$.
If the asymmetries are magnetic in origin, $\nu^\ell_B$ must be overestimated by a factor of $\gtrsim6$.
It is plausible that order-unity uncertainties in Equation \ref{bcrit} (for $B_{r,\mathrm{crit}}$) and $a^{\ell=1}_{^-\!1\,0\,^+\!1}$ may account for this.
However, the measured asymmetries require very different field strengths for different radial orders.
This suggests either that the magnetic field falls off quickly in the outer layers or that the asymmetries have a non-magnetic origin.
In the latter case, the asymmetries may instead originate from GD 358's drifting mode frequencies.
This frequency wandering does not keep the asymmetry constant over time \citep[cf. Figure 9 of][]{bischoff2019gd358}.
\new{Notably, after the \textit{sforzando} event of August 1996 during which the $k=8$ and $k=9$ modes briefly dominated the light curve, the asymmetries of the $k=8$ and $k=9$ triplets were permanently modified \citep{provencal20092006,montgomery2010evidence}.}
We therefore suspect that the measured asymmetries result from wandering mode frequencies rather than magnetic shifts.

\new{Curiously, \citet{montgomery2010evidence} show that the $k=12$ triplet in GD 358 exhibits signatures of oblique pulsation, which occurs when the Lorentz force exceeds the Coriolis force in strength and misaligns the pulsations from the rotation axis \citep[the $k=12$ triplet was not observed by the Whole Earth Telescope May 1990 run data examined in this work;][]{winget1994whole}.
In the case of oblique pulsation, the magnetic field still shifts mode frequencies by $\sim\nu^\ell_B$.
However, contrary to the prediction in Section \ref{sectfrequencyshifts} (which assumes rotationally aligned pulsations), the magnetic field does not simply produce an asymmetric rotational multiplet from the perspective of an inertial observer.
Instead, each mode manifests as multiple peaks in the power spectrum \citep{loi2021topology}.
In this case, the asymmetry-based upper bounds $B_{r,\mathrm{shift}}$ in this work and \citet{winget1994whole} should be replaced by a more elaborate fit to an oblique pulsator model.
The parameters extracted by \citet{montgomery2010evidence} in this way are consistent with a magnetic field strength $B_r\sim40\,\mathrm{kG}\sim10B_{\mathrm{crit}}$, placing their result at tension with our assumptions regarding the fundamental g-mode suppression physics (Section \ref{sectmodesuppression}).}

\citet{bagnulo2022multiple} conduct a volume-limited spectropolarimetric survey of nearby WDs, finding a dearth of magnetic fields in young ($\lesssim2\,\mathrm{Gyr}$), canonical-mass WDs down to their detection limit (a few kilogauss).
These spectropolarimetric measurements are shown in Figure \ref{fig:total_field_summary}.
The low ($\lesssim10\,\mathrm{kG}$) magnetic field limits we place on most of our carbon--oxygen WDs are consistent with this finding.
In many cases, our seismic limits appear to be much more constraining, entailing magnetic fields less than $\sim100\,\mathrm{G}$ in several WDs with long-period pulsation modes.

\new{In addition to placing upper limits on magnetism in individual WDs, the observation of (non-suppressed) pulsations can constrain the distribution of magnetic fields in the WD population as a whole.
If the DAVs included in this work are representative, the high purity of the instability strip \citep[e.g.,][]{gianninas2005toward} suggests that most DA WDs at instability-strip temperatures lack magnetic fields $\gtrsim10\,\mathrm{kG}$.}

\subsection{ELMVs: probing the stripped cores of red giants?}

\begin{figure}
    \centering
    \includegraphics[width=0.46\textwidth]{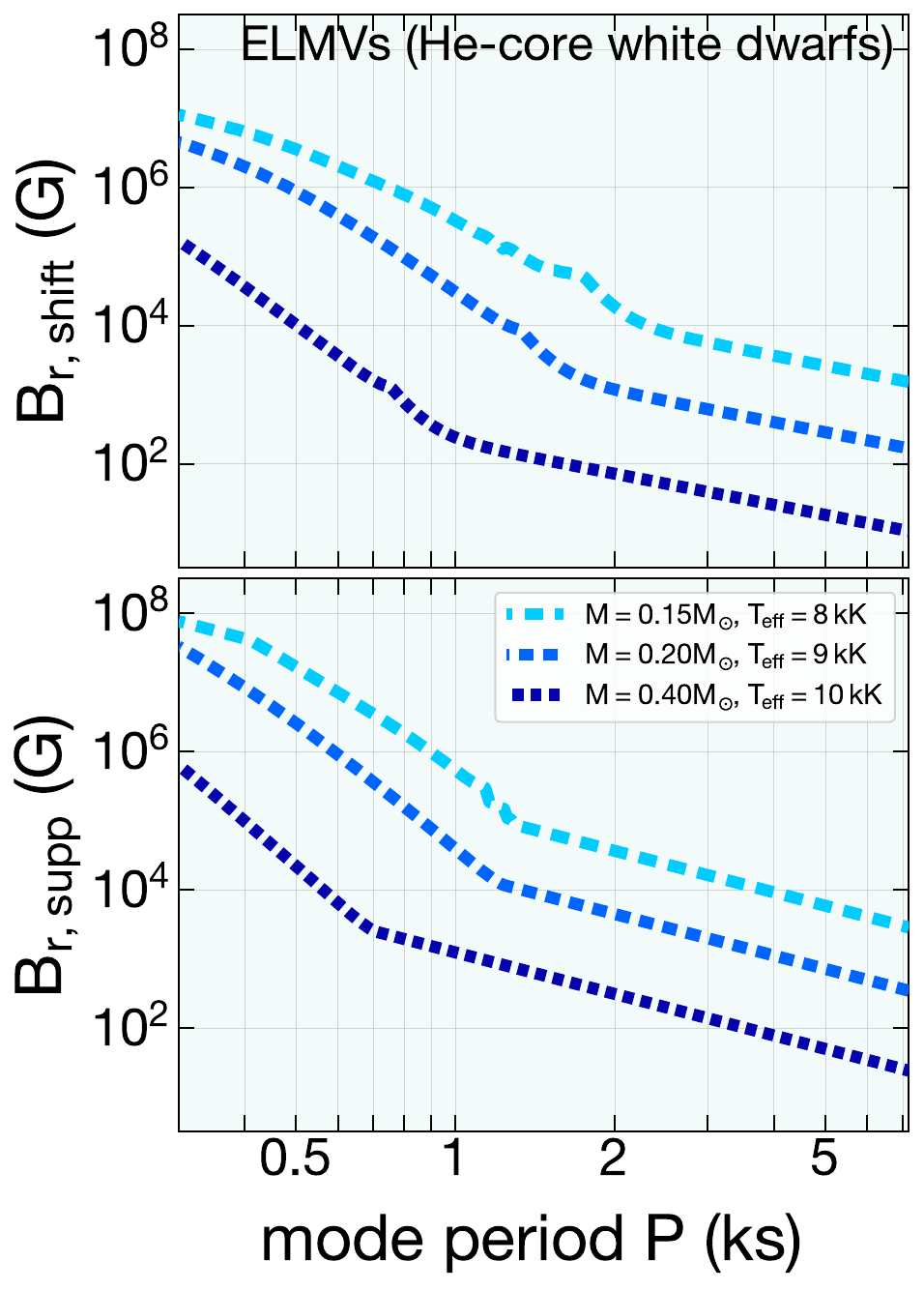}
    \caption{Model seismic magnetic field sensitivities $B_{r,\mathrm{shift}}$ and $B_{r,\mathrm{supp}}$ for $0.15M_\odot$, $0.20M_\odot$, and $0.40M_\odot$ helium-core WD pulsators on the DAV instability strip.
    These curves have been calculated for $\ell=1$ and $\nu^{\ell=1}_B=0.1\,\mu\mathrm{Hz}$.
    }
    \label{fig:show_constraining_power_elmv}
\end{figure}

Unlike carbon--oxygen WDs, helium-core WDs are exclusively produced by binary stellar evolution.
These low-mass ($M\lesssim0.5M_\odot$) WDs are the leftover cores of red giants whose envelopes have been stripped by a companion or ejected during a common-envelope event.
To date, roughly 20 helium-core WDs are known to pulsate \citep{hermes2012sdss,hermes2013discovery,hermes2013new,kilic2015psr, gianninas2016discovery,bell2017pruning,bell2018mcdonald,kilic2018refined,pelisoli2018sda,pelisoli2019sda,wang2020discovery}.
These ``extremely low-mass variables,'' or ELMVs, are all are located on a low-$\log g$ extension of the DAV instability strip \citep{gianninas2015elm,tremblay20153d}.
As the direct descendants of the same red giants for which seismic magnetometry is already possible, helium-core WDs likely inherit the field strengths of the cores of their progenitors.
Figure \ref{fig:show_constraining_power_elmv} shows the magnetic fields to which helium-core WD pulsators are sensitive.

\begin{figure}
    \centering
    \includegraphics[width=0.46\textwidth]{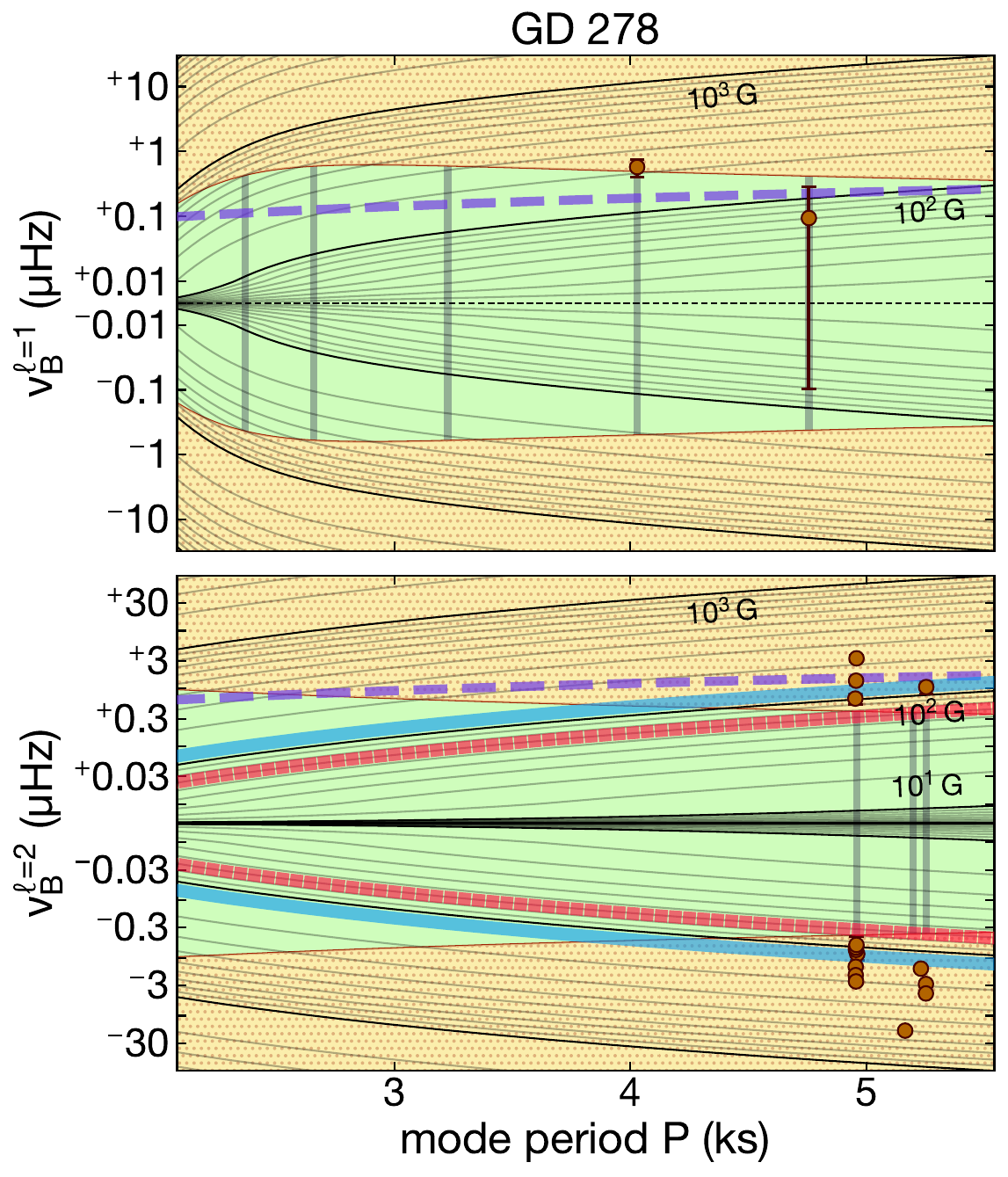}
    \caption{Same as Figure \ref{fig:fieldplot_DAV}, but for the ELMV GD 278 \citep{lopez2021discovery,calcaferro2023exploring}.
    The \textit{dashed purple curves} show the predicted asymmetries from second-order rotational effects on the dipole asymmetry $\Delta^{\ell=2}_{^-\!1\,0\,^+\!1}$ and a typical quadrupole asymmetry $\Delta^{\ell=2}_{^-\!1\,0\,^+\!2}$.}
    \label{fig:fieldplot_ELMV}
\end{figure}

We place an upper bound on GD 278, a low-mass ($M\approx0.19M_\odot$) WD which pulsates in low-frequency gravity modes \citep[$P\simeq2400$--$6700\,\mathrm{s}$;][]{lopez2021discovery}.
\new{To date, GD 278 is the only known ELMV for which rotational splittings have been measured.}
Assuming the mode identifications determined by \citet{calcaferro2023exploring}, we find that the magnetic field in GD 278 is no larger than $B_{r,\mathrm{supp}}\sim70\,\mathrm{G}$ (Figure \ref{fig:fieldplot_ELMV}).
This field is much lower than those inferred for dipole-suppressed red giants \citep[$\approx20\%$ of all red giants;][]{stello2016prevalence}, and smaller than the field expected to be generated by a core convective dynamo on the main sequence \citep{cantiello2016asteroseismic}.

GD 278's asymmetries are often close to or within the forbidden region, i.e., within the \textit{yellow dotted region} in Figure \ref{fig:fieldplot_ELMV}).
As discussed for GD 358, a magnetic origin for these asymmetries may still be possible if some order-unity prefactors are misestimated.
The relatively fast rotation period $P_{\mathrm{rot}}\approx10\,\mathrm{hr}$ \citep{lopez2021discovery} corresponds to $P/P_{\mathrm{rot}}\sim10\%$ for the observed modes, suggesting that second-order rotational effects may play a role (shown as the \textit{dashed purple curves} in Figure \ref{fig:fieldplot_ELMV}). 
Second-order rotational effects are expected to produce asymmetries $\Delta^{\ell=1}_{^-\!1\,0\,^+\!1}=P/20P_{\mathrm{rot}}^2\simeq0.2\,\mu\mathrm{Hz}$ in the two complete observed triplets at $P\approx4030\,\mathrm{s}$ and $P\approx4760\,\mathrm{s}$ \citep{dziembowski1992effects}.
The observed asymmetries ($\Delta^{\ell=1}_{^-\!1\,0\,^+\!1}=(0.44\pm0.06)\,\mu\mathrm{Hz}$ and $\Delta^{\ell=1}_{^-\!1\,0\,^+\!1}=(0.07\pm0.07)\,\mu\mathrm{Hz}$, respectively) are similar in scale to (though at significant tension with) those predicted by second-order rotational effects.
Further, we find that observed asymmetries do not consistently match second-order rotational asymmetries in sign.
We also curiously find that observed quadrupole asymmetries are often lower than the expected second-order asymmetries by factors $\sim3$--$10$ relative to predicted values \citep{dziembowski1992effects}.
The ultimate origin of asymmetries in GD 278 are puzzling.

It is possible that GD 278's progenitor was a lower-mass ($M\lesssim1.2M_\odot$) main-sequence star below the Kraft break.
Such stars have radiative cores, and are thus not expected to generate strong fields during the main sequence.
If this is the case, it suggests that, in the absence of a convective core, the core field of a main-sequence star can be extremely low.
In a $1M_\odot$ MESA model, mass coordinates $\leq0.19M_\odot$ are enclosed by a radius $R\approx0.1R_\odot$ during the main-sequence.
This decreases to $\approx0.025R_\odot$ during the red giant branch, once the helium core has grown to $\approx0.19R_\odot$.
Assuming flux conservation, this implies that the core magnetic field of GD 278's progenitor was no larger than $\simeq4\,\mathrm{G}$ on the main sequence.
Alternatively, our upper bound may challenge the assumption of magnetic flux conversation between evolutionary stages.

At a time of rapid progress in red giant magnetoasteroseismology, ELMVs may help test our understanding of magnetism in stellar cores as well as the essential pulsation physics itself.
Since the near-surface WD fields to which seismology is sensitive can also be probed by spectroscopy or spectropolarimetry, comparison between the field strengths inferred by these techniques and seismology may serve as a valuable consistency check.
\new{While thorough spectroscopic surveys of ELMs have been conducted \citep[e.g.,][]{brown2010elm,brown2011elm,brown2020elm,wang2022extremely,kosakowski2023elm}, we are not presently aware of any detections of Zeeman splitting in low-mass WDs (although \citealt{pichardo2023candidate} and \citealt{marcano2024second} report photometric variability in two candidate ELM WDs in the globular cluster NGC 6397 that might arise from magnetic spots).}




\subsection{Caveats}

The behavior of high-frequency pulsations under magnetic fields should be investigated in more detail.
While we make a rudimentary correction for some non-asymptotic effects in this work, our correction leaves out some additional effects.
Non-asymptotic modes may produce significant displacements in evanescent regions, which are ignored by our integral estimate for $\mathscr{I}$ in Equation \ref{curlyIasympt}.
Further, as described in Appendix \ref{sectnonasymptotic}, our results correct for non-asymptotic eigenfunctions in the calculation of $\mathscr{I}$.
However, the derivation of Equation \ref{numag} (which writes $\nu^\ell_B$ in terms of $\mathscr{I}$) still assumes certain terms scaling as $\sim2\pi\nu/N$ are negligible, including effects of the horizontal field components and the radial displacement in the calculation of the mode inertia.
Our estimates of magnetic effects also assume the mode wavelength is smaller than the scale height $H$ on which quantities such as $\rho$ and $B$ change.
For the long-period modes shown in Figure \ref{fig:fancy_profile_main}, the mode wavelengths are comparable to $H$ just below the convective zone, making magnetic effects on modes more difficult to calculate.
The effect of relaxing these assumptions should be investigated in more detail.

Our analysis also required basic assumptions about the dimensionless asymmetry parameters $a^\ell_{m_1\,m_2\,m_3}$, as well as the prefactor which appears in Equation \ref{bcrit} for $B_{r,\mathrm{crit}}$.
Both of these are order-unity functions of the field geometry \citep{lecoanet2017conversion,rui2023gravity,lecoanet2022asteroseismic} and the rotation rate, if it is high enough \citep{rui2024asteroseismic}.
This introduces an uncertain order-unity factor to our field estimates \citep[e.g.,][]{das2024unveiling}.
However, dipole/quadrupole asymmetries may fail more catastrophically when faced with, e.g., small-scale magnetic structures that cannot be resolved by $\ell=1$ or $\ell=2$ modes well enough to produce significant rotational multiplet asymmetries.

We also assume that the frequencies of each multiplet component have not been affected by nonlinear resonant mode coupling.
However, WD pulsation modes occasionally show large, nonsecular frequency changes \citep[e.g.,][]{hermes2013newpdot,dalessio2013periodic}.
In fact, some multiplet components within a pulsating WD have shown correlated, periodic frequency changes, a signature of resonant mode coupling \citep{zong2016amplitude}.
Such frequency changes could produce non-magnetic multiplet asymmetries.

Finally, this work stitches together a perturbative theory in $(B_r/B_{r,\mathrm{crit}})^2$ for $B_r<B_{r,\mathrm{crit}}$ with the assumption of total suppression when $B_r>B_{r,\mathrm{crit}}$.
In reality, perturbation theory is no longer valid once $B_r$ is a large fraction of $B_{r,\mathrm{crit}}$.
\citet{rui2024asteroseismic} show, for example, that the inappropriate application of perturbation theory overestimates the field strength in the case of a centered dipolar field aligned with the rotation axis.
Along similar lines, strong-enough fields in slow-enough rotators are expected to align pulsations with the magnetic axis, dramatically increasing the complexity of a rotational multiplet in the power spectrum.
These strong magnetic effects should be more thoroughly investigated.

\section{Summary and future directions} \label{summaryfuture}

We summarize our conclusions in this work as follows:

\begin{enumerate}
    \item WD pulsations can be used as a highly sensitive probe of the radial components of their near-surface magnetic fields.
    Magnetic fields are expected to asymmetrize rotational multiplets at intermediate field strengths $B_r\lesssim B_{r,\mathrm{crit}}$ and suppress g-mode propagation when $B_r>B_{r,\mathrm{crit}}$.
    
    \item Seismology is most sensitive to magnetic fields in pulsating WDs with lower-frequency modes and thinner surface convection zones.

    \item We place upper limits on the magnetic field in $20$ DAV and $3$ DBV carbon--oxygen WDs.
    Most upper limits in carbon--oxygen WDs lie within roughly $\sim1$--$10\,\mathrm{kG}$, but they can be as low as a \textit{few to tens of gauss} when low-frequency modes ($\gtrsim800\,\mathrm{s}$) are not observed to be suppressed.
    Seismology may help uncover the as-of-yet unknown formation mechanism of strong magnetic fields in WDs at low effective temperatures.

    \item We also place an upper limit on the near-surface magnetic field $\simeq70\,\mathrm{G}$ in the ELMV GD 278 (helium-core) WD.
    As stripped red giants, ELMVs may present a valuable opportunity to test our understanding of red giant magnetism as well as the fundamental pulsation physics.
    
\end{enumerate}

\new{Following results from red giant asteroseismology, this work assumes that strong magnetic fields $B_r>B_{r,\mathrm{crit}}$ strongly suppress gravity waves \citep{fuller2015asteroseismology,stello2016prevalence,cantiello2016asteroseismic}.
However, despite some informative theoretical work \citep{lecoanet2017conversion,loi2017torsional,loi2018effects,loi2020effect,lecoanet2022asteroseismic,rui2023gravity}, some details of the damping mechanism are not fully understood.
If pulsations can instead persist in the presence of a strong magnetic field, our suppression-based upper bounds on the field strength $B_{r,\mathrm{supp}}$ no longer apply, although our asymmetry-based upper bounds $B_{r,\mathrm{shift}}$ (which are grounded on more certain physics and are typically of similar order) are still valid.}

\new{Although definitive evidence of magnetic g-mode suppression remains largely elusive, the massive WD SDSS J1529+2928 may be an example of magnetic g-mode suppression acting in a white dwarf.
Despite lying within the DAV instability strip, SDSS J1529 has not been observed to pulsate.
Furthermore, although spectroscopy excludes a magnetic field $\gtrsim70\,\mathrm{kG}$, SDSS J1529 exhibits variability consistent with a magnetic spot \citep{kilic2015dark}.
SDSS J1529 may therefore be an illustrative example of magnetic suppression of oscillations, either by damping the gravity waves magnetically (the mechanism invoked in this work) or by obstructing the convection involved in driving the pulsations \citep[not considered in this work, although see, e.g.,][]{tremblay2015evolution}.}

Seismic WD magnetometry affords many further possibilities.
Pulsating WDs are a highly diverse class which also include pre-WDs \citep[GW Virs/DOVs/PNNVs, e.g.,][]{winget1991asteroseismology,sowicka2023gw}, pre-ELMVs \citep[e.g.,][]{maxted2013multi}, and pulsating accretors in cataclysmic variables \citep[GW Lib, e.g.,][]{van2004non}.
Additionally, several pulsating ultramassive ($M\gtrsim1.1M_\odot$) DAV WDs are known \citep{hermes2013discoveryumv,curd2017four,kanaan2005whole,rowan2019detections,vincent2020searching,kilic2023wd} which, unlike the WDs considered in this work, have already begun to crystallize on the DAV instability strip.
This is expected to affect their pulsations \citep{kanaan2005whole,nitta2015constraining,de2019pulsation}, possibly allowing them to help uncover the relationship between their crystallization and magnetization.
Finally, some hot subdwarfs are known to pulsate in gravity modes \citep{pablo2012seismic,telting2014low,baran2016subsynchronously,baran2019k2,silvotti2019high,sanjayan2022pulsating,silvotti2022filling,ma2023amplitude}, which are likely sensitive to their near-core fields.
Seismology may contribute to the solution of an open problem regarding the dearth of observed \new{magnetic fields in} hot subdwarf\new{s} \citep{dorsch2022discovery,pelisoli2022discovery,pakmor2024large,dorsch2024discovery}.

\vspace{1em}
\noindent We thank Sivan Ginzburg for helpful discussions and comments on the manuscript\new{, as well as the anonymous referee for their useful suggestions}.
We are grateful for support from the United States--Israel Binational Science Foundation through grant BSF-2022175.
N.Z.R. acknowledges support from the National Science Foundation Graduate Research Fellowship under Grant No. DGE‐1745301.

\vspace{5mm}
\facilities{None}
\software{MESA \citep{paxton2010modules,paxton2013modules,paxton2015modules,paxton2018modules}, NumPy \citep{oliphant2006guide}, SciPy \citep{virtanen2020scipy}, AstroPy \citep{robitaille2013astropy}, Matplotlib \citep{Hunter:2007,the_matplotlib_development_team_2024_13308876}}

\bibliography{sample631}{}
\bibliographystyle{aasjournal}

\appendix

\section{Magnetic formalism} \label{formalism}

The asymmetry-based magnetic field bounds placed in this work require concrete assumptions about the asymmetry parameters $a^\ell_{m_1\,m_2\,m_3}$.
This Appendix derives the dipole and quadrupole asymmetry parameters assumed in this study.

\subsection{Magnetic frequency shifts of g-mode pulsations}

In the most common formalism \citep{bugnet2021magnetic,li2022magnetic,das2024unveiling}, the Coriolis and Lorentz forces are assumed to be weak and degenerate perturbation theory is applied.
The incompressible and asymptotic approximations (appropriate for high-radial order g modes) are additionally assumed.
Furthermore, it is assumed either that the magnetic field is axisymmetric about the rotation axis or that the effect of the Coriolis force on the waves is much stronger than that of the Lorentz force.
In both cases, the eigenfunctions are aligned with the rotation axis, and the frequency shifts are only sensitive to azimuthal averages of $B_r^2$, i.e.,
\begin{equation}
    \langle B_r^2\rangle_\phi = \int^{2\pi}_0\frac{\mathrm{d}\phi}{2\pi}\,B_r^2\mathrm{,}
\end{equation}

\noindent although the field's dependence on the colatitude $\mu=\cos\theta$ still matters.

In the inertial frame, the frequency shifts $\delta\nu$ of the dipole modes are given by Equations 32 and 33 of \citet{li2022magnetic}:

\begin{subequations} \label{dipole1}
    \begin{gather}
        \delta\nu^{\ell=1}_{m=0} = \mathcal{B}^{\ell=1}\left[\frac{3}{2}(1-\mu^2)\right] \\
        \delta\nu^{\ell=1}_{m=\pm 1} = \pm\frac{1}{2}\frac{\langle\Omega\rangle_g}{2\pi} + \mathcal{B}^{\ell=1}\left[\frac{3}{4}(1+\mu^2)\right]\mathrm{,}
    \end{gather}
\end{subequations}

\noindent and the same for the quadrupole modes are given by Appendix C.1 of \citet{das2024unveiling}:

\begin{subequations} \label{quadrupole1}
    \begin{gather}
        \delta\nu^{\ell=2}_{m=0} = \mathcal{B}^{\ell=2}\left[\frac{15}{2}(\mu^2-\mu^4)\right] \\
        \delta\nu^{\ell=2}_{m=\pm1} = \pm\frac{5}{6}\frac{\langle\Omega\rangle_g}{2\pi} + \mathcal{B}^{\ell=2}\left[\frac{5}{4}(1-3\mu^2+4\mu^4)\right] \\
        \delta\nu^{\ell=2}_{m=\pm2} = \pm\frac{5}{3}\frac{\langle\Omega\rangle_g}{2\pi} + \mathcal{B}^{\ell=2}\left[\frac{5}{4}(1-\mu^4)\right]\mathrm{,}
    \end{gather}
\end{subequations}

\noindent where the relevant integral average operation $\mathcal{B}^\ell[\,\cdot\,]$ is given under the present assumptions by
\begin{equation} \label{helperoperator}
    \mathcal{B}^\ell[f(\mu)] \approx \nu^\ell_B\int_{\mathcal{R}^\ell_{\;\nu}}\mathrm{d}r\,K(r)\int_{S^2}\frac{\mathrm{d}\Omega}{4\pi}\,B_r^2\,f(\mu)\mathrm{.}
\end{equation}

If the Lorentz force is instead comparable to or stronger than the Coriolis force, the frequency shifts it causes are still at the same order of magnitude as $\nu^\ell_B$ (as in Equation \ref{numag}), although the dimensionless prefactors are given by the solution of a more complicated matrix problem, and multiplets generally possess more than $2\ell+1$ peaks in the inertial frame (see the discussion in the Supplementary Information of \citealt{li2022magnetic}).
This does not seem to be the case in any of the WDs we analyze, since all of their (possibly incomplete) rotational multiplets are broadly recognizable and lack significant extra frequency peaks.

\subsection{Multiplet asymmetries due to magnetism}


When $K(r)$ is sharply peaked and the magnetic frequency shifts are only sensitive to a geometrically thin radial shell (which is often true, especially in the WD case), the radial integral in Equation \ref{helperoperator} becomes independent of the angular one, and we can speak of a single horizontal dependence of $B_r$.
In other words, we can approximately take
\begin{equation}
    B_r(r,\mu,\phi) = A(r)\,\psi(\mu,\phi)\mathrm{,}
\end{equation}

\noindent where we normalize the horizontal dependence of $B_r$ to $\psi=\psi(\mu,\phi)$ such that $\int_{S^2}\psi^2\,\sin\theta\,\mathrm{d}\theta\,\mathrm{d}\phi=1$.
We define the azimuthal average $\langle\psi^2\rangle_\phi$ as
\begin{equation} \label{azavg}
    \langle\psi^2\rangle_\phi \equiv \int^{2\pi}_0\frac{\mathrm{d}\phi}{2\pi}\,\psi^2\mathrm{.}
\end{equation}

Under this condition, \citet{li2022magnetic} show that the dipole asymmetry parameter is given by
\begin{equation} \label{asym1}
    a^{\ell=1}_{^-\!1\,0\,^+\!1} = \int^{+1}_{-1}\langle\psi^2\rangle_\phi\,\frac{1}{2}\left(3\mu^2 - 1\right)\mathrm{d}\mu\mathrm{,}
\end{equation}

\noindent and encodes some information about the geometry of the field.

We generalize the parameter $a^{\ell=1}_{^-\!1\,0\,^+\!1}$ to describe the asymmetry between any three modes within the same multiplet, including those involving higher-degree ($\ell>1$) modes such as quadrupole modes ($\ell=2$).
We start by noticing that the linear combination of frequencies in Equation \ref{dipolefreqcombo} is useful because it depends on neither the unperturbed frequencies $\nu^{(0)}$ (which are the same for all modes within a multiplet) nor the rotational splitting (which is proportional to $m$).
In other words, for three modes with equal $k$ and $\ell$ but distinct azimuthal orders $m_1$, $m_2$, and $m_3$, maximally ``useful'' linear combinations $\Delta^\ell_{m_1\,m_2\,m_3}$ of measured frequencies of the form
\begin{equation}
    \Delta^\ell_{m_1\,m_2\,m_3} = c_1\nu^\ell_{m_1} + c_2\nu^\ell_{m_2} + c_3\nu^\ell_{m_3}\mathrm{.}
\end{equation}

\noindent should satisfy $c_1+c_2+c_3=0$ and $m_1c_1+m_2c_2+m_3c_3=0$ to cancel out $\nu^{(0)}$ and $\langle\Omega\rangle_g$, respectively.

For concreteness, for the quadrupole modes, we choose the coefficients $c_1$, $c_2$, and $c_3$ to be small integers.
There are $_{2\ell+1}C_3=10$ distinct useful linear combinations:
\begin{subequations} \label{allquadrupoles}
    \begin{align}
        \Delta^{\ell=2}_{^-\!2\,^-\!1\,0} &= \nu^{\ell=2}_{m=^-\!2} -2\,\nu^{\ell=2}_{m=^-\!1} + \nu^{\ell=2}_{m=0} \\
        \Delta^{\ell=2}_{^-\!2\,^-\!1\,^+\!1} &= 2\,\nu^{\ell=2}_{m=^-\!2} -3\,\nu^{\ell=2}_{m=^-\!1} + \nu^{\ell=2}_{m=^+\!1} \\
        \Delta^{\ell=2}_{^-\!2\,^-\!1\,^+\!2} &= 3\,\nu^{\ell=2}_{m=^-\!2} -4\,\nu^{\ell=2}_{m=^-\!1} + \nu^{\ell=2}_{m=^+\!2} \\
        \Delta^{\ell=2}_{^-\!2\,0\,^+\!1} &= \nu^{\ell=2}_{m=^-\!2} -3\,\nu^{\ell=2}_{m=0} + 2\,\nu^{\ell=2}_{m=^+\!1} \\
        \Delta^{\ell=2}_{^-\!2\,0\,^+\!2} &= \nu^{\ell=2}_{m=^-\!2} -2\,\nu^{\ell=2}_{m=0} + \nu^{\ell=2}_{m=^+\!2} \\
        \Delta^{\ell=2}_{^-\!2\,^+\!1\,^+\!2} &= \nu^{\ell=2}_{m=^-\!2} -4\,\nu^{\ell=2}_{m=^+\!1} + 3\,\nu^{\ell=2}_{m=^+\!2} \\
        \Delta^{\ell=2}_{^-\!1\,0\,^+\!1} &= \nu^{\ell=2}_{m=^-\!1} -2\,\nu^{\ell=2}_{m=0} + \nu^{\ell=2}_{m=^+\!1} \\
        \Delta^{\ell=2}_{^-\!1\,0\,^+\!2} &= 2\,\nu^{\ell=2}_{m=^-\!1} -3\,\nu^{\ell=2}_{m=0} + \nu^{\ell=2}_{m=^+\!2} \\
        \Delta^{\ell=2}_{^-\!1\,^+\!1\,^+\!2} &= \nu^{\ell=2}_{m=^-\!1} -3\,\nu^{\ell=2}_{m=^+\!1} + 2\,\nu^{\ell=2}_{m=^+\!2} \\
        \Delta^{\ell=2}_{0\,^+\!1\,^+\!2} &= \nu^{\ell=2}_{m=0} -2\,\nu^{\ell=2}_{m=^+\!1} + \nu^{\ell=2}_{m=^+\!2}\mathrm{,}
    \end{align}
\end{subequations}

Although only $2\ell-1=3$ of these contain independent information, observed multiplets are often incomplete, restricting which of these $10$ asymmetry parameters can be calculated.
In general, whether a given mode is observable depends on the excitation mechanism as well as viewing angle (see \citealt{gizon2003determining} and \citealt{das2024unveiling} for further discussion).
WD modes are also not generally in energy equipartition, and even modes within the same multiplet can be excited to very different amplitudes \citep{hermes2015insights}.

Following the normalization convention of \citet{li2022magnetic} and \citet{das2024unveiling}, these linear combinations are related to dimensionless asymmetry parameters $a^\ell_{m_1\,m_2\,m_3}$ using Equations \ref{dipoleasym} and \ref{quadrupoleasym}.
Using Equation \ref{azavg}, we rewrite the operator $\mathcal{B}^\ell[\,\cdot\,]$ as
\begin{equation}
    \mathcal{B}^\ell[f(\mu)] \approx \nu^\ell_B\int^{+1}_{-1}\,\langle\psi^2\rangle_\phi\,f(\mu)\,\mathrm{d}\mu\mathrm{.}
\end{equation}

Equations \ref{quadrupole1} for the quadrupole frequency shifts then imply that

\begin{subequations} \label{asym2}
    \begin{align}
        \begin{split}
             a^{\ell=2}_{^-\!2\,^-\!1\,0} &= a^{\ell=2}_{0\,^+\!1\,^+\!2} \\
             = &\int^{+1}_{-1}\langle\psi^2\rangle_\phi\,\frac{1}{4}\left(-15\mu^{4} + 12 \mu^{2} - 1\right)\mathrm{d}\mu \\
        \end{split} \\
        \begin{split}
             a^{\ell=2}_{^-\!2\,^-\!1\,^+\!1} &= a^{\ell=2}_{^-\!2\,^-\!1\,^+\!2} = a^{\ell=2}_{^-\!2\,^+\!1\,^+\!2} = a^{\ell=2}_{^-\!1\,^+\!1\,^+\!2} \\
             = &\int^{+1}_{-1}\langle\psi^2\rangle_\phi\,\frac{1}{2}\left(-5 \mu^{4} + 3 \mu^{2}\right)\mathrm{d}\mu \\
        \end{split} \\
        \begin{split}
            a^{\ell=2}_{^-\!2\,0\,^+\!1} &= a^{\ell=2}_{^-\!1\,0\,^+\!2} \\
            = &\int^{+1}_{-1}\langle\psi^2\rangle_\phi\,\frac{1}{4}\left(25 \mu^{4} - 24 \mu^{2} + 3\right)\mathrm{d}\mu \\
        \end{split} \\
        \begin{split}
            a^{\ell=2}_{^-\!2\,0\,^+\!2} = \int^{+1}_{-1}\langle\psi^2\rangle_\phi\,\frac{1}{2}\left(5 \mu^{4} - 6 \mu^{2} + 1\right)\mathrm{d}\mu \\
        \end{split} \\
        \begin{split}
            a^{\ell=2}_{^-\!1\,0\,^+\!1} = \int^{+1}_{-1}\langle\psi^2\rangle_\phi\,\frac{1}{2}\left(10 \mu^{4} - 9 \mu^{2} + 1\right)\mathrm{d}\mu\mathrm{.}
        \end{split}
    \end{align}
\end{subequations}

Curiously, five groups of dimensionless asymmetry parameters in Equations \ref{asym2} are constrained to be equal.
This is not a field geometry-dependent fact, but rather a result of the assumption that rotation fixes the preferred direction respected by the modes.
It is also distinct from the linear dependence of the linear combinations of frequencies in Equations \ref{allquadrupoles}, although they are self-consistent.
For example, Equations \ref{allquadrupoles} and \ref{quadrupoleasym} easily show that $\Delta^{\ell=2}_{^-\!1\,0\,^+\!1}+2\Delta^{\ell=2}_{0\,^+\!1\,^+\!2}=\Delta^{\ell=2}_{^-\!1\,^+\!1\,^+\!2}$, a relationship which is obeyed by the expressions in Equations \ref{asym2}.

On the one hand, this implies that not all of the modes in a quintuplet are required to be measured to extract all of the information the quintuplet encodes.
Conversely, checking that these linear combinations of quadrupole modes in fact obey these relationships may be a useful test in determining whether observed asymmetries are in fact magnetic in origin (under the present assumptions).

\subsection{Asymmetry parameters for an inclined dipole} \label{inclined}

For a centered dipole magnetic field with some obliquity angle $\beta$ relative to the rotation axis, the normalized horizontal dependence of $B_r$ is given by
\begin{equation}
    \psi(\mu,\phi) = \frac{1}{2}\sqrt{\frac{3}{\pi}}(\cos\beta\cos\theta + \sin\beta\sin\theta\cos\phi)\mathrm{,}
\end{equation}
cf. \citet{mathis2023asymmetries} and \citet{das2024unveiling}.

Averaging $\psi^2$ over $\phi$ gives
\begin{equation} \label{important}
    \langle\psi^2\rangle_\phi = \frac{3}{2}P_2(\cos\beta)\mu^2 + \frac{3}{4}\sin^2\beta
\end{equation}

\noindent where $P_2(x)=(3x^2-1)/2$ is a Legendre polynomial.

The form of Equation \ref{important} is highly instructive.
It writes $\langle\psi^2\rangle_\phi$ as two terms, the second of which is a constant over the star and therefore shifts all modes equally (i.e., cannot introduce asymmetries).
Since all asymmetry parameters only depend on the field through latitudinal averages over $\langle\psi^2\rangle_\phi$, \textit{every} asymmetry parameter defined in Equations \ref{asym1} and \ref{asym2} must be proportional to $P_2(\cos\beta)$ (and therefore to each other, in ratios which are independent of $\beta$).
Moreover, \textit{all} asymmetry parameters of any $\ell$ (including $\ell>2$) must vanish for all $\beta$ at some critical obliquity $\beta=\arccos(-1/3)/2$ where $P_2(\cos\beta)$ vanishes (this has been noticed in special cases by \citealt{mathis2023asymmetries} and \citealt{das2024unveiling}).
This appears to be a special property of the dipole geometry, where the relevant component of $\langle\psi^2\rangle_\phi$ depends on $\beta$ and $\mu$ only in a ``disentangled'' way.

Evaluating Equations \ref{asym1} and \ref{asym2}, we have
\begin{equation}
    a^{\ell=1}_{^-\!1\,0\,^+\!1} = \frac{2}{5}P_2(\cos\beta)
\end{equation}

\noindent for $\ell=1$, and

\begin{subequations} \label{armyoflegendre}
    \begin{gather}
        a^{\ell=2}_{^-\!2\,^-\!1\,0} = a^{\ell=2}_{^-\!1\,0\,^+\!1} = a^{\ell=2}_{0\,^+\!1\,^+\!2} = - \frac{2}{35}P_2(\cos\beta) \\
        a^{\ell=2}_{^-\!2\,^-\!1\,^+\!1} = a^{\ell=2}_{^-\!2\,0\,^+\!1} = a^{\ell=2}_{^-\!1\,0\,^+\!2} = a^{\ell=2}_{^-\!1\,^+\!1\,^+\!2} = - \frac{6}{35}P_2(\cos\beta) \\
        a^{\ell=2}_{^-\!2\,^-\!1\,^+\!2} = a^{\ell=2}_{^-\!2\,^+\!1\,^+\!2} = - \frac{12}{35}P_2(\cos\beta) \\
        a^{\ell=2}_{^-\!2\,0\,^+\!2} = - \frac{8}{35}P_2(\cos\beta)
    \end{gather}
\end{subequations}

\noindent for $\ell=2$.

If all possible magnetic axes are equally probable, $P_2(\cos\beta)$ has a root-mean-square value
\begin{equation}
    \begin{split}
        P_2(\cos\beta)_{\mathrm{rms}} &= \sqrt{\frac{1}{2}\int^{+1}_{-1}P_2(\cos\beta)^2\,\mathrm{d}(\cos\beta)} \\
        &= \sqrt{\frac{2}{5}} \approx 0.63
    \end{split}
\end{equation}

We adopt this root-mean-square value as the fiducial value of $P_2(\cos\beta)$ when modeling asymmetry parameters in this study.

\section{Correction for non-asymptotic effects} \label{sectnonasymptotic}

The exposition in Section \ref{theory} makes extensive use of the asymptotic approximation and related assumptions.
For example, the expressions in Section \ref{theory} require incompressibility, as well as the condition that the radial wavelength is small compared both to the horizontal wavelength $\lambda_h/2\pi=r/\sqrt{\ell(\ell+1)}$ and the pressure scale height.
Both assumptions are challenged for WD pulsations, which are usually localized to the outer edge of the g-mode cavity near the surface of the star, and are low-radial order at the high-frequency end.
Non-asymptotic effects may be particularly important for the magnetism, whose effects are especially confined to the outer turning point of the g-mode cavity.

\begin{figure}
    \centering
    \includegraphics[width=0.6\textwidth]{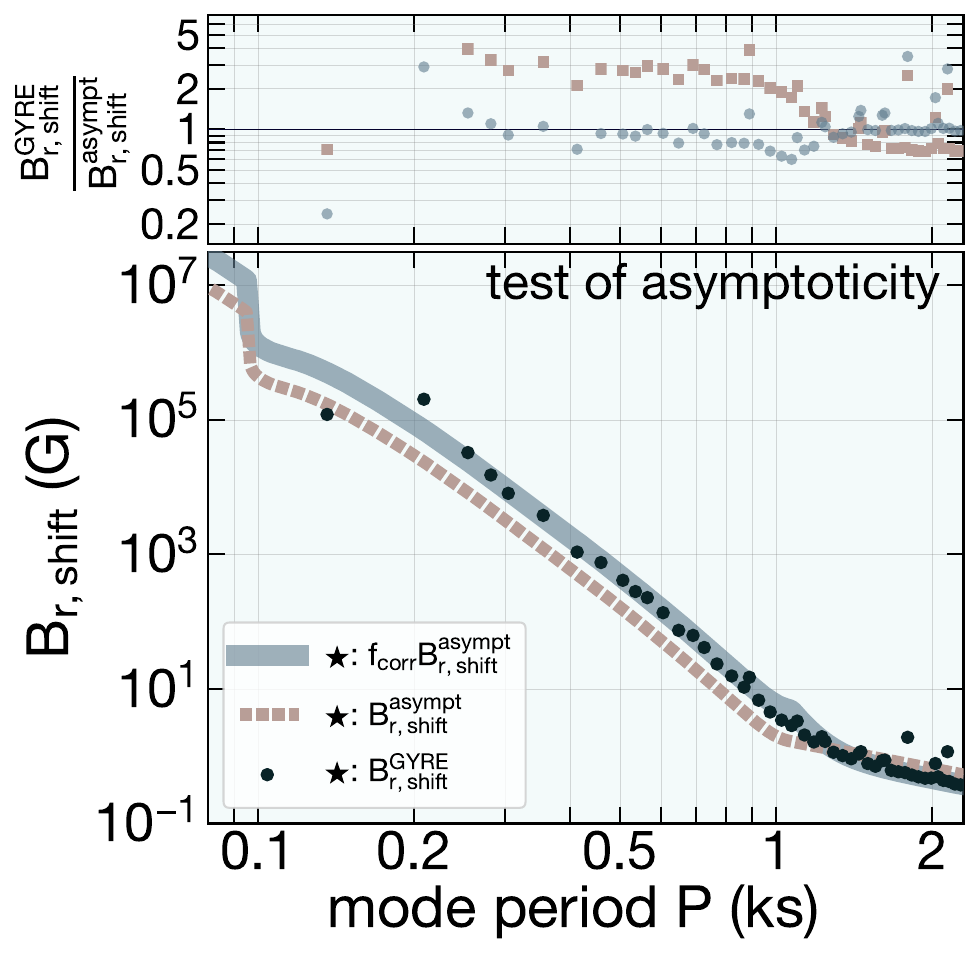}
    \caption{Model seismic magnetic field sensitivity $B_{r,\mathrm{shift}}$, with or without the non-asymptotic correction in Equation \ref{fcorr} applied.
    \textit{Black points} denote the non-asymptotic result using eigenfunctions calculated by GYRE (Equation \ref{nonasympt}).
    The \textit{top panel} shows the ratio of the \new{GYRE result to the} asymptotic estimate (with or without correction) to the non-asymptotic result.
    This figure has been extended to longer periods to emphasize the behavior of $N$-limited modes.
    The symbol $\bigstar$ denotes the same choice of parameters as in Figure \ref{fig:show_constraining_power_polish}.}
    \label{fig:show_constraining_power_nonasympt}
\end{figure}

To investigate the effect of relaxing these assumptions, we numerically solve for the adiabatic oscillation modes using version 7.2.1 of the GYRE code \citep{townsend2013gyre}.
GYRE computes both mode frequencies and fluid perturbations in the absence of assumptions about the size of the density scale height or perturbations to the gravitational potential.
Equation \ref{curlyIasympt} gives an asymptotic estimate of $\mathscr{I}$.
We then compute a non-asymptotic estimate $\mathscr{I}$ using the mode periods $P$ and horizontal fluid displacements $\xi_h$ from GYRE \citep[cf. Equation 40 in][]{li2022magnetic}:
\begin{equation} \label{nonasympt}
    \mathscr{I} = \frac{4\pi^2}{\ell(\ell+1)}P^{-2}\frac{\int_{\mathcal{R}^\ell_{\;\nu}}[\partial_r(r\xi_h)]^2\mathrm{d}r}{\int_{\mathcal{R}^\ell_{\;\nu}}\xi_h^2\rho r^2\,\mathrm{d}r}\mathrm{.}
\end{equation}

The derivation of Equation \ref{nonasympt} applies assumptions such as incompressibility and approximately radial wavenumber, and it ignores a surface term contribution which has been found to make a small difference \citep{jones1989possibility}.
Nevertheless, it stops short of substituting in the full asymptotic expression for $\xi_h$ \citep[see the Appendix of][]{li2022magnetic}.
GYRE does not assume that the radial wavenumber is large relative to the scale height, nor does it assume the Cowling approximation.
In the asymptotic limit (taking $\xi_h\propto\rho^{-1/2}r^{-3/2}N^{1/2}\sin\Phi$ with $k_r=\partial_r\Phi=\sqrt{\ell(\ell+1)}N/\omega r$), we recover Equation \ref{curlyIasympt}.
However, $\mathscr{I}$ may differ from its value in Equation \ref{curlyIasympt} for various reasons, including corrections to the outer turning point and domination of the integrals in Equation \ref{nonasympt} by only a single wavelength.

We bundle non-asymptotic effects into a simple, ad hoc correction factor $f_{\mathrm{corr}}$, defined such that the inferred field $B_{r,\mathrm{shift}}$ from the asymmetries is related to its asymptotic estimate by
\begin{equation} \label{fcorr}
    f_{\mathrm{corr}} = \frac{B_{r,\mathrm{shift}}}{B^{\mathrm{asympt}}_{r,\mathrm{shift}}} = \sqrt{\frac{\mathscr{I}^{\mathrm{asympt}}}{\mathscr{I}}}
\end{equation}

\noindent where superscript ``asympt'' denotes the application of Equation \ref{curlyIasympt} for calculating $\mathscr{I}$.

We find that $f_{\mathrm{corr}}$ is approximately described by
\begin{equation}
    f_{\mathrm{corr}} = f_N + (f_S - f_N)\left(\frac{2\pi}{PS_\ell}\right)_{\mathrm{out}}^{10}
\end{equation}

\noindent where $f_{N}=1/\sqrt{2}$ and $f_{S_\ell}=3$, respectively, and $\left(2\pi/PS_\ell\right)_{\mathrm{out}}$ is evaluated at the outer turning point of the mode.
This form is chosen to set $f_{\mathrm{corr}}=f_{N}$ for $N$-limited modes and $f_{\mathrm{corr}}=f_{S_\ell}$ for $S_\ell$-limited modes, with a fast but smooth transition in between the regimes enforced by an arbitrary but steep power index ($10$).
Figure \ref{fig:show_constraining_power_nonasympt} shows that there is good agreement between $B^{\mathrm{GYRE}}_{r,\mathrm{shift}}$ (computed using GYRE eigenfunctions) and $B_{r,\mathrm{shift}}=f_{\mathrm{corr}}B^{\mathrm{asympt}}_{r,\mathrm{shift}}$.
Future work should more thoroughly investigate the impact of non-asymptotic effects on seismic magnetic field measurements.


\end{document}